\pdfoutput=1

\documentclass[11pt]{article}

\usepackage[final]{acl}

\usepackage{times}
\usepackage{latexsym}

\usepackage[T1]{fontenc}

\usepackage[utf8]{inputenc}

\usepackage{microtype}

\usepackage{inconsolata}

\usepackage{graphicx}
\usepackage{pifont}
\usepackage{graphicx}
\usepackage{amsmath}
\usepackage{amssymb}
\usepackage{hyperref}[colorlinks,linkcolor=blue]
\usepackage{float} 
\usepackage{subfigure} 
\usepackage{booktabs,makecell, multirow, tabularx}
\usepackage{algorithm}
\usepackage{algorithmic}  

\usepackage[normalem]{ulem}
\usepackage{color}
\useunder{\uline}{\ul}{}
\newcommand{\Rmnum}[1]{\uppercase\expandafter{\romannumeral #1}} 
 
\title{LSSF: Safety Alignment for Large Language Models through Low-Rank Safety Subspace Fusion}

\author{
 \textbf{Guanghao Zhou\textsuperscript{1}\thanks{Completed during internship at Ant Group}},
 \textbf{Panjia Qiu\textsuperscript{1}\footnotemark[\value{footnote}]},
 \textbf{Cen Chen \textsuperscript{1}\thanks{Corresponding author}},
 \textbf{Hongyu Li\textsuperscript{2}},
\\
 \textbf{Mingyuan Chu\textsuperscript{2}},
 \textbf{Xin Zhang\textsuperscript{2}},
 \textbf{Jun Zhou\textsuperscript{2}},
\\
 \textsuperscript{1}East China Normal University 
 \textsuperscript{2}Ant Group
\\
\texttt{\{ghzhou, panjiaqiu\}@stu.ecnu.edu.cn, cenchen@dase.ecnu.edu.cn} \\
\texttt{zhiyuan.lhy@antgroup.com, jszjg1991@gmail.com} \\ 
\texttt{evan.zx@ant-intl.com, jun.zhoujun@antgroup.com}}

\begin{document}
\maketitle
\begin{abstract}

The safety mechanisms of large language models (LLMs) exhibit notable fragility, as even fine-tuning on datasets without harmful content may still undermine their safety capabilities. Meanwhile, existing safety alignment methods predominantly rely on the fine-tuning process, which inadvertently leads to the increased complexity and computational resources required. To address these issues, we introduce LSSF, a novel safety re-alignment framework with \underline{L}ow-Rank \underline{S}afety \underline{S}ubspace \underline{F}usion. Our proposed method exploits the low-rank characteristics of safety information in LLMs by constructing a low-rank projection matrix to extract the principal components of safety vectors. Notably, this projection matrix represents the low-rank safety subspace of the LLMs, which we have observed to remain stable during fine-tuning process and is isolated from the model's general capabilities. These principal components are used to effectively restore safety alignment when combined with fine-tuned LLMs through linear arithmetic. Additionally, to account for the varying encoding densities of safety information across different layers of LLMs, we propose a novel metric called safety singular value entropy. This metric quantifies the encoding density and allows for the dynamic computation of the safety-critical rank for each safety vector. Extensive experiments demonstrate that our proposed post-hoc alignment method can effectively restore the safety alignment of fine-tuned models with minimal impact on their performance in downstream tasks.
\end{abstract}
                                          
\section{Introduction}
In recent years, as the capabilities of large language models (LLMs) have improved significantly \cite{achiam2023gpt,llama3modelcard}, a growing amount of research has focused on enhancing their safety to prevent unsafe responses that conflict with human values \cite{Christiano2017DeepRL, Yuan2023RRHFRR}. Numerous studies have revealed that aligned LLMs exhibit surprising safety vulnerabilities after fine-tuning \cite{Qi2023FinetuningAL, zhan-etal-2024-removing, fan2025trustworthiness}. 
The safety of these models can be significantly compromised when updated with  a small amount of maliciously crafted or even benign data.
To address this issue, existing studies \cite{Zong2024SafetyFA, Huang2024LazySA} have primarily focused on ensuring model safety and consistency by aligning the model during the fine-tuning phase. 
However, these methods not only increase the complexity of the training process and require additional computational resources, but they also potentially inhibit the model's general capabilities. 
Recent work \cite{Ilharco2022EditingMW} demonstrates that adding or subtracting task vectors, i.e., directional vectors corresponding to specific tasks within a model, can enhance or reduce the model's performance on those tasks.
Building on this, RESTA \cite{bhardwaj-etal-2024-language} introduces a post-hoc alignment method that restores the safety of compromised models by performing an arithmetic combination with safety vectors. Nonetheless, as the safety vectors contain elements that inhibit general capabilities, the integrated model may experience an inevitable reduction in its general abilities.

\begin{figure}[t]
    \centering
    \includegraphics[width=\columnwidth]{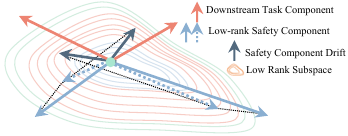}
    \caption{Illustration of using low-rank safety principal components to restore model safety alignment, where all safety components share the same low-rank subspace.}
    \label{fig:low-rank}
\end{figure}

Recent studies \cite{Sun2023ASA, Wei2024AssessingTB} have indicated that the safety 
regions in LLMs are isolated and sparse at the rank level, 
distinct from the directions of the models' general capabilities. 
Our experiments in Section \ref{sec:low-rank-exp} 
also confirm that the safety drift directions of safety vectors likewise exhibit low-rank properties and share a common low-rank safety subspace with LLMs.

Based on the above insights, we 
introduce LSSF, a novel safety re-alignment framework with \underline{L}ow-Rank \underline{S}afety \underline{S}ubspace \underline{F}usion.
Specifically, we perform low-rank orthogonal matrix decomposition on the activations of the safety-aligned LLMs and construct a projection matrix to extract the low-rank principal components of the corresponding safety vectors.
As illustrated in Figure \ref{fig:low-rank}, when the critical safety directions of a fine-tuned model drift, we can effectively rectify this deviation by applying linear arithmetic to the low-rank principal components of the safety vectors. 

Moreover, we propose a safety singular value entropy information density quantification method inspired by Shannon entropy \cite{Shannon1948Comm}. 
Previous work on assessing brittleness of safety alignment~\cite{Wei2024AssessingTB} reveals that different linear layers in LLMs encode safety and general capabilities to varying degrees. This highlights the necessity of determining the pruning rank of the corresponding safety vector based 
on the density of safety information encoded in the weight matrix.
Our singular value entropy considers both the absolute magnitudes of the singular values and their relative distribution. By analyzing the proportion of singular value entropy, we can effectively control information loss when truncating the rank.
Extensive experiments on Qwen2.5-7B-Instruct \cite{qwen2.5} and Llama3.1-8B-Instruct \cite{llama3modelcard} demonstrate that our proposed LSSF can restore safety alignment with minimal impact on the downstream task performance of their fine-tuned models.

Our contributions are summarized as follows:
\begin{itemize}
\item We proposed the utilization of a projection matrix to extract the low-rank principal components of the safety vector, enabling the safety realignment of the fine-tuned LLMs within this low-rank subspace.
\item We proposed a novel safety singular value entropy-based information density quantification method that effectively assesses the safety information encoding density within the linear layer and assists in the determination of the appropriate pruning rank for the safety vector.
\item  We performed comprehensive experiments on various LLMs, which demonstrated that our method can effectively restores their safety alignment without significantly compromising the downstream task performance.
\end{itemize}

\section{Related Work}

\textbf{Safety Realignment.} 
Pre-trained LLMs are typically enhanced for specific downstream tasks through a process known as supervised fine-tuning, which often involves full fine-tuning \cite{howard-ruder-2018-universal} and parameter-efficient fine-tuning \cite{Hu2021LoRALA, ben-zaken-etal-2022-bitfit}.
Even LLMs with strong initial safety alignment can be manipulated to produce harmful content during the fine-tuning process \cite{Bianchi2023SafetyTunedLL, He2024WhatII}.  
Some studies \cite{Dai2023SafeRS, Huang2024LazySA, Bianchi2023SafetyTunedLL} focus on ensuring safety realignment during the fine-tuning of LLMs, which undoubtedly increases the complexity of this process. 
RESTA \cite{bhardwaj-etal-2024-language} employs the direct arithmetic combination of safety vectors to the weights of fine-tuned models. Since safety vectors include general capability-suppressing components, there is a certain impact on the model's performance on downstream tasks.

\begin{figure*}[htbp]
    \centering
    \includegraphics[width=\linewidth]{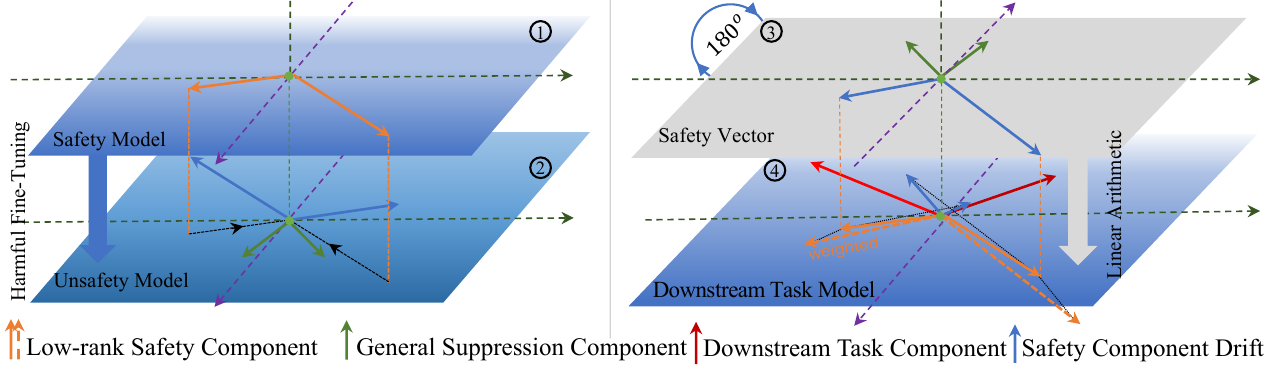}
    \caption{Overview of our safety re-alignment framework. Step \Rmnum{1}: \ding{172} $\rightarrow$ \ding{173}, obtain an unsafe model through unsafe fine-tuning. Step \Rmnum{2}: \ding{173} $\rightarrow$ \ding{174}, invert the delta parameters to derive the safety vector. Step \Rmnum{3}: \ding{174} $\rightarrow$ \ding{175}, construct a low-rank projection matrix to extract the safety principal components of the safety vector and perform linear arithmetic with the downstream fine-tuned model to restore its safety alignment.}
    \label{fig:main}
\end{figure*}

\noindent \textbf{Low-Rank Compression and Pruning.} Unstructured pruning techniques aim to establish criteria based on weight magnitude, activations, or network gradients to remove individual weights from the network \cite{cao-etal-2021-low, guo-etal-2021-parameter, zhang-etal-2024-loraprune}. Low-rank compression techniques are similar to structured pruning methods, focusing on identifying important structured sub-networks  \cite{Sun2023ASA, Frantar2023SparseGPTML, Wang2024SVDLLMTS}.  ActSVD \cite{Wei2024AssessingTB} extracts the safety-critical rank of LLMs through singular value decomposition (SVD) of stacked activations and demonstrates its low-rank nature. We extend the low-rank pruning of LLMs to safety vectors to extract principal components of the corresponding low-rank secure subspace.

\noindent \textbf{Model fusion.} Current fusion methods for LLMs generally fall into three categories: geometric \cite{SLERP}, pruning \cite{Yadav2023TIESMergingRI}, and arithmetic \cite{xiao-etal-2024-lm}. As a geometry-based approach, Model Stock \cite{Jang2024ModelSA} considers the geometric properties in the weight space. Pruning-based methods such as Breadcrumbs \cite{Davari2023ModelBS} and DARE \cite{Yu2023LanguageMA} eliminate interference among multiple models by removing redundant parameters. Arithmetic-based methods include Average Merging \cite{Wortsman2022ModelSA} and Task Arithmetic \cite{bhardwaj-etal-2024-language}. The former merges models by averaging parameters, while the latter introduces task vectors and uses scaling terms to adjust the importance of different models. We extend task arithmetic to safety fine-tuning tasks and employ low-rank pruning methods to mitigate its impact on the LLMs' downstream task performance.

\section{Methodology}
Our objective is to restore the safety of LLMs through post-hoc alignment.  
Figure \ref{fig:main} illustrates our safety re-alignment framework. Starting from an unsafe model derived via harmful fine-tuning, we compute the inverted safety vector and extract its low-rank safety components. These are then integrated into the downstream task model through linear arithmetic to restore safety alignment.
In Section \ref{subsec:safety-vector}, we introduce the safety vectors of LLMs, Sections \ref{subsec:low-rank-pruning} and \ref{subsec:entropy-quan} provide a detailed explanation of how to extract low-rank safety components and Section \ref{subsec:lin-ari} explains how task arithmetic restores the safety of fine-tuned models.

\subsection{Safety Vector}
\label{subsec:safety-vector}

Safety vector is derived from the delta parameters when transitioning from the unsafe base model to the safety-aligned model, formulated as:
\begin{align}
    \boldsymbol{\theta}_\text{safe} = \boldsymbol{\theta}_\text{unsafe} + \boldsymbol{\delta}_\text{safe},
\end{align}
where $\boldsymbol{\theta}_\text{safe}$ denotes the parameters of the safety-aligned model, while $\boldsymbol{\theta}_\text{unsafe}$ refers to the parameters of the unsafe model and $\boldsymbol{\delta}_\text{safe}$ represents the safety vector obtained through the alignment process. However, compromising safety guardrails is significantly easier than safety alignment, as the former only requires fine-tuning on a small amount of toxic data. As shown in Figure \ref{fig:main} Step \Rmnum{1}, we use the toxic dataset $\mathcal{D}_\text{unsafe} = \{(x_i, y_i) \mid i = 1, \ldots, N\}$ to perform supervised fine-tuning on $\boldsymbol{\theta}_\text{safe}$ 
to obtain the inverse safety vector $-\boldsymbol{\delta}_{\text{safe}}$,where $x_i$ represents harmful queries, and $y_i$ denotes affirmative responses to these harmful queries.

As shown in Figure \ref{fig:main} \ding{173}, $-\boldsymbol{\delta}_{\text{safe}}$ consists of two main components. The first is the drift of the low-rank safety component in the opposite direction of safety, which can be extracted through low-rank decomposition due to its low-rank nature. The second is the general suppression component, which impairs the model's performance on general tasks.

\subsection{Low-rank Orthogonal Decomposition}
\label{subsec:low-rank-pruning}

Motivated by ActSVD \cite{Wei2024AssessingTB}, 
we perform singular value decomposition on the linear layer activations of $\boldsymbol{\theta}_{\text{base}}$ and use the left singular vectors to construct a low-rank projection matrix.
First, we construct a calibration dataset $\mathcal{D}_{\text{anchor}}=\{(x'_i, y'_i)|i=1, \cdots, N'\}$, where $x'_i$ represents harmful queries and $y'_i$ represents safe negative responses. For any linear layer weight matrix $W \in \mathbb{R}^{d_{\text{out}} \times d_{\text{in}}}$, we obtain the corresponding input matrix $\widehat{X} \in \mathbb{R}^{d_{\text{in}} \times n}$ from $\mathcal{D}_{\text{anchor}}$. The objective of the low-rank decomposition of $\boldsymbol{\theta}_\text{base}$ is to achieve a low-rank approximation of $W$ while maintaining its safety performance. Specifically, we seek a rank-$r$ low-rank matrix $\widehat{W}$ that minimizes the Frobenius norm of the output changes, as described by the following formula:
\begin{align}
    \widehat{W}=\underset{\operatorname{rank(}\widehat{W})\leq r}{\operatorname*{\operatorname*{\operatorname*{\operatorname*{\arg\min}}}}}\|W\widehat{X}-\widehat{W}\widehat{X}\|_{F}^{2},
\end{align}
where the optimal low-rank matrix $\widehat{W}$ shares the same low-rank subspace as the safe vector $\delta_\text{safe}$. The proof is provided in Appendix \ref{app-sec:low-rank-subspace}. 

To eliminate the sensitivity differences of neurons in the linear layer to different texts in the anchor dataset, we normalize the activation matrix $Z = W\widehat{X} = [z_{ij}]_{d_{in} \times n}$ as follows:
\begin{align}
\label{equ:tilde-z}
    \tilde{z}_{ij} = \frac{z_{ij} - \mu_{j}}{\delta_j},
\end{align}
where $z_{i,j}$ represents the activation value at $(i,j)$, $\mu_j$ and $\delta_j$ denote the mean and standard deviation of each column, respectively.
We perform a low-rank matrix decomposition on the standardized activation matrix $\tilde{Z}$ using SVD:
\begin{align}
\label{equ:svd}
    USV^{\top} \approx \tilde{Z},
\end{align}
where $S = \text{diag}(\sigma_1, \sigma_2, \cdots, \sigma_n)$ is the singular value diagonal matrix and $U \in \mathbb{R}^{d_{\text{out}} \times n}$ is an orthogonal matrix composed of the top $n$ left singular vectors. Due to the slow speed of SVD for large matrices, we follow the method from \citet{Halko2009FindingSW}, which uses a randomized algorithm to efficiently calculate approximate solutions of SVD.

\subsection{Low-Rank Projection Matrix}
\label{subsec:entropy-quan}

For matrix $\tilde{Z}$ in Equation \ref{equ:tilde-z}, the square of the Frobenius norm can be expressed as $\|\tilde{Z}\|^2_F = \sum_{i,j} |\tilde{z}_{i,j}|^2$. According to the SVD, we have $\|\tilde{Z}\|_F^2 = \sum_{i=1}^n \sigma_i^2$, where $\sigma_i$ represents the $i$-th singular value, the proof can be found in Appendix \ref{app-sec:singular-value-entropy}. This indicates that the energy of the matrix $\tilde{Z}$ is equal to the sum of the squares of its singular values, which can be interpreted as a measure of the overall complexity or information content of the matrix. 
Therefore, we quantify each principal component's contribution to the total information content using its squared singular value, defined as:
\begin{align}
    p_i = \frac{\sigma_i^2}{\sum_{j=1}^n \sigma_j^2},
\end{align}
where $p_i$ represents the information contribution of the $i$-th principal component. Singular value entropy is used to evaluate the complexity and information content of matrix, taking into account not only the absolute magnitudes of the squared singular values but also their relative distribution. The formula for calculating singular value entropy is as follows:
\begin{align}
    H_{\rho}  = -\sum^{\rho}_{i=1} \frac{\sigma^2_i}{\sum_{j=1}^n \sigma_j^2}\log \left( \frac{\sigma^2_i}{\sum^n_{j=1}\sigma^2_j}\right),
\end{align}
where $H_{\rho}$ represents the entropy of the singular values of the top $\rho$ ranks. We use the information retention threshold $\eta$ to determine the rank preserved by the orthogonal projection:
\begin{align}
    \frac{H_{r}}{H_n} > \eta,
\end{align}
where $r$ represents the rank to be retained. Accordingly, we construct a low-rank projection matrix:
\begin{align}
    P^{(r)} = U^{(r)}\left( U^{(r)} \right)^{\top},
\end{align}
where $P^{(r)}$ denotes the orthogonal projection matrix onto the $r$ most significant left singular subspaces, where $\text{rank}(P^{(r)}) = r$. The proof is provided in Appendix \ref{app-sec:SVD}.

\noindent To balance the utility and safety of LLMs, we apply the scaling factor $\alpha$ to the singular vectors to enhance or diminish their drift in the corresponding safety directions. The scaling factor $\alpha_i$ corresponding to the singular vector $u_i \in U^{(r)}$ is calculated as follows:
\begin{align}
    \alpha_i = 1 + (\alpha_1 - 1) \times \frac{\sigma_i - \sigma_r}{\sigma_1 - \sigma_r},
\end{align}
where $\alpha_1$ represents the weighting factor for the singular vector corresponding to the largest singular value, and subsequent weights decreasing proportionally. This formulation effectively enhances the weights of singular vectors corresponding to larger singular values in a proportional manner. The weighted singular vectors are:
\begin{align}
    U'^{(r)} = \left(u_1, \cdots, u_r\right) \begin{pmatrix}
\alpha_1 &    &    \\
 &    \ddots &  \\
   &    & \alpha_r
\end{pmatrix}.
\end{align}
\indent Therefore, the final low-rank projection matrix corresponding to the  safety vector is:
\begin{align}
   P'^{(r)}=U'^{(r)}\left( U'^{(r)} \right)^\top. 
\end{align}

\subsection{Linear Arithmetic}
\label{subsec:lin-ari}

Let $\boldsymbol{\theta}_{\text{DST}}$ represent the model parameters of $\boldsymbol{\theta}_{\text{safe}}$ after fine-tuning for downstream task datasets. The fine-tuning process may compromise the safety guardrails of LLMs and it could be expressed as:
\begin{align}
\boldsymbol{\theta}_{\text{DST}} &= \boldsymbol{\theta}_{\text{safe}} + \boldsymbol{\delta}_{\text{DST}} \notag \\
&= \boldsymbol{\theta}_{\text{safe}} + \boldsymbol{\tau}_{\text{DST}} - \boldsymbol{\tau}_{\text{safe}} + \hat{\boldsymbol{\tau}}_{\text{DST}},
\end{align}
were $\boldsymbol{\delta}_{\text{DST}}$ represents the delta parameters obtained from the SFT of $\boldsymbol{\theta}_{\text{safe}}$. We decompose $\boldsymbol{\delta}_{\text{DST}}$ into the desired downstream task direction offset $\boldsymbol{\tau}_{\text{DST}}$, an undesired offset in the safety direction $-\boldsymbol{\tau}_{\text{safe}}$ and a redundant shift in other directions $\hat{\boldsymbol{\tau}}_{\text{DST}}$.
Our objective is to add a low-rank safety component $\boldsymbol{\tau}'_{\text{safe}}$ to $\boldsymbol{\delta}_{\text{DST}}$ to counteract the $-\boldsymbol{\tau}_{\text{safe}}$ offset while minimizing impacts on shifts in other directions. Our objective could be expressed as:
\begin{align}
\boldsymbol{\theta}_{\text{DST}}' &= \boldsymbol{\theta}_{\text{safe}} + \boldsymbol{\delta}_{\text{DST}} + \alpha P'^{(r)}\boldsymbol{\delta}_{\text{safe}} \notag \\
&= \boldsymbol{\theta}_{\text{safe}} + \boldsymbol{\tau}_{\text{DST}} - \boldsymbol{\tau}_{\text{safe}} + \alpha \boldsymbol{\tau}'_{\text{safe}} + \ \hat{\boldsymbol{\tau}} \notag \\
&\approx \boldsymbol{\delta} + \boldsymbol{\tau}_{\text{DST}} + \hat{\boldsymbol{\tau}},
\end{align}
where $\hat{\boldsymbol{\tau}}$ denotes the drift in the redundant direction, which does not significantly affect $\boldsymbol{\tau}_{\text{DST}}$. Additionally, $\boldsymbol{\tau}_{\text{safe}}$ and $\boldsymbol{\tau}'_{\text{safe}}$ share the same low-rank subspace, as detailed in Section \ref{subsec:entropy-quan}.

\section{Experiment}

\subsubsection{Models under testing}
We perform experiments on two LLMs: Qwen2.5-7B-Instruct (Qwen2.5) \cite{qwen2.5} and Llama3.1-8B-Instruct (Llama3.1) \cite{llama3modelcard}. Due to their strong safety and instruction-following capabilities, we adopt them as safety models and base models for downstream fine-tuning.
In Appendix \ref{app-sec: robu-paras}, we selected Qwen2.5-3B-Instruct and Qwen2.5-14B-Instruct to validate the robustness of our method across models with varying parameters.

\subsubsection{Baselines}
We classify baseline methods into two categories: alignment during supervised fine-tuning (SFT) and post-hoc alignment. The baselines include: Non-Alignment SFT (NA-SFT), which does not enforce safety alignment; VLGuard \cite{Zong2024SafetyFA} and Lisa \cite{Huang2024LazySA}, which implement safe alignment during the fine-tuning process; RESTA \cite{bhardwaj-etal-2024-language}, which applies safety alignment after fine-tuning. For detailed information on the baselines and specific experimental settings, please refer to Appendix \ref{app-sec:baseline}.

\subsubsection{Datasets}

\textbf{Calibration Dataset.}
We construct a calibration dataset to obtain the safety-critical ranks of the safety model. To ensure the harmfulness of the queries, we selected samples from PKU-SafeRLHF \cite{ji2024pku} where both the \textit{accept} and \textit{reject} responses were labeled as \textit{unsafe}. As demonstrated in Appendix \ref{app-sec: robu-num-calibra}, our method is robust to the number of calibration data samples. Consistent with \cite{Wei2024AssessingTB}, we ultimately chose a total of 128 such samples. These harmful queries are then input into the safety model to collect safe refusal responses. Based on the pairs of harmful queries and safe responses, we construct a calibration dataset.

\noindent \textbf{Downstream Fine-Tuning Dataset.} 
To better demonstrate the effectiveness of our method in downstream task fine-tuning experiments, we established two distinct scenarios. The first scenario involves LoRA fine-tuning, where we utilize AG's News and Yahoo Answers \cite{Zhang2015CharacterlevelCN} for multi-class classification tasks. The second scenario is full fine-tuning, which performs poorly on LoRA since it not only requires knowledge infusion but also focuses on dialogue-based question answering. In this scenario, we emphasize the medical knowledge question-answering dataset\footnote{https://github.com/Toyhom/Chinese-medical-dialogue-data}, specifically designed for question-answering and text generation tasks.
For detailed construction of the SFT datasets, please refer to Appendix \ref{app-sec:dataset}.

\subsubsection{Safety Vector Calculation}
We calculate safety vectors by determining the offset between the aligned models and its unaligned counterparts. To construct a harmful fine-tuning dataset, we randomly select 500 harmful queries and their corresponding harmful responses labeled as \textit{unsafe} from PKU-SafeRLHF \cite{ji2024pku}. The dataset is subsequently utilized to perform SFT for 3 epochs on Qwen2.5 and Llama3.1 to get new models that compromise safety guardrails.

\begin{table*}[t!]
\centering
\resizebox{\textwidth}{!}{
\begin{tabular}{lcccccccc}
\toprule
\multirow{2}{*}{Methods} & \multicolumn{4}{c}{Qwen2.5-7B-Instruct} & \multicolumn{4}{c}{Llama3.1-8B-Instruct} \\ \cline{2-9} 
& ACC $\uparrow$  & AdvBench $\uparrow$ & HarmfulQA $\uparrow$ & CATQA $\uparrow$ & ACC $\uparrow$   & AdvBench $\uparrow$ & HarmfulQA $\uparrow$ & CATQA $\uparrow$ \\ 
\midrule
NA-SFT  & \underline{0.91} & 0.09 & 0.50 & 0.19 & \textbf{0.86} & 0.04 & 0.49 & 0.12   \\
VLGuard & \textbf{0.92} & 0.78 & \underline{0.54} & 0.13 & \underline{0.85} & 0.92 & 0.56 & 0.21   \\
RESTA   & \textbf{0.92} & \underline{0.99} & \textbf{0.98} & 0.92 & 0.69 & 0.51 & 0.91 & 0.72   \\
Lisa    & 0.90 & \underline{0.99} & \textbf{0.98} & \textbf{0.94} & 0.80 & \underline{0.96} & \underline{0.92}  & \underline{0.96}   \\
Ours   & \textbf{0.92} & \textbf{1.00} & \textbf{0.98} & \underline{0.93} & \underline{0.85} & \textbf{1.00}  & \textbf{1.00} & \textbf{1.00}   \\ \bottomrule
\end{tabular}}
\caption{Performance of different safety alignment methods in the AG's News LoRA SFT Task. \textbf{\textit{ACC}} represents the classification accuracy. \textbf{\textit{AdvBench}}, \textbf{\textit{HarmfulQA}}, and \textbf{\textit{CATQA}} denote refusal rates for the corresponding datasets.}
\label{tab:mainAlignResults-AGNews}
\end{table*}

\begin{table*}[t!]
\centering
\resizebox{\textwidth}{!}{
\begin{tabular}{lcccccccc}
\toprule
\multirow{2}{*}{Methods} & \multicolumn{4}{c}{Qwen2.5-7B-Instruct} & \multicolumn{4}{c}{Llama3.1-8B-Instruct} \\ \cline{2-9} 
 & ACC $\uparrow$   & AdvBench $\uparrow$ & HarmfulQA $\uparrow$ & CATQA $\uparrow$ & ACC $\uparrow$   & AdvBench $\uparrow$  & HarmfulQA $\uparrow$ & CATQA $\uparrow$ \\ \midrule
NA-SFT  & \textbf{0.68} & 0.20 & 0.52 & 0.12 & \underline{0.63} & 0.04 & 0.49 & 0.12   \\
VLGuard & \textbf{0.68} & 0.96 & 0.58 & 0.21 & \textbf{0.64}    & 0.87 & 0.51 & 0.15   \\
RESTA   & 0.63          & 0.94 & 0.92 & 0.81 & 0.25             & 0.33 & 0.86 & 0.62   \\
Lisa    & \underline{0.67} & \underline{0.98} & \underline{0.97} & \underline{0.93} & 0.45 & \underline{0.89} & \underline{0.97} & \underline{0.89}   \\
Ours   & \textbf{0.68} & \textbf{1.00} & \textbf{0.99} & \textbf{0.99} & \textbf{0.64} & \textbf{1.00} & \textbf{0.99} & \textbf{0.99} \\ \bottomrule
\end{tabular}}
\caption{Performance of different safety alignment methods in the Yahoo Answers LoRA SFT Task . \textbf{\textit{ACC}} represents the classification accuracy. \textbf{\textit{AdvBench}}, \textbf{\textit{HarmfulQA}}, and \textbf{\textit{CATQA}} denote refusal rates for the corresponding datasets.}
\label{tab:mainAlignResults-Yahoo}
\end{table*}

\begin{table*}[t!]
\centering
\resizebox{\textwidth}{!}{
\begin{tabular}{lcccccccccc}
\toprule
\multicolumn{1}{c}{\multirow{2}{*}{Method}} & \multicolumn{5}{c}{Qwen2.5-7B-Instruct}        & \multicolumn{5}{c}{Llama3.1-8B-Instruct}       \\ \cline{2-11} 
\multicolumn{1}{c}{}      
& BLUE $\uparrow$ & Rouge-L $\uparrow$ & AdvBench $\uparrow$ & HarmfulQA $\uparrow$ & CATQA $\uparrow$ & BLUE $\uparrow$ & Rouge-L $\uparrow$ & AdvBench $\uparrow$ & HarmfulQA $\uparrow$ & CATQA $\uparrow$ \\ \midrule
NA-SFT   & \textbf{0.44} & \textbf{0.52} & 0.21 & 0.54 & 0.15 & \textbf{0.44} & \textbf{0.53} & 0.11 & 0.49 & 0.12  \\
VLGuard  & \textbf{0.44} & \textbf{0.52} & \underline{0.88} & 0.62 & \underline{0.21} & \underline{0.42} & \underline{0.50} & \underline{0.98} & \underline{0.62} & 0.34  \\
RESTA    & 0.34 & 0.43 & \textbf{0.99} & \underline{0.91} & \textbf{0.95} & 0.18 & 0.27 & 0.91 & \textbf{0.99} & 0.97  \\
Lisa     & 0.31 & 0.41 & \textbf{0.99} & \underline{0.91} & \textbf{0.95} & 0.21 & 0.31 & 0.94 & \textbf{0.99} & \underline{0.98}  \\
Ours     & \underline{0.42} & \underline{0.50} & \textbf{0.99} & \textbf{0.94} & \textbf{0.95} & \underline{0.42} & \underline{0.50} & \textbf{0.99} & \textbf{0.99} & \textbf{0.99}  \\ \bottomrule
\end{tabular}}
\caption{Performance of different safety alignment methods in the Medical QA Full SFT Task. \textbf{\textit{BLUE}} and \textbf{\textit{Rouge-L}} assess the consistency between the generated text
and the reference text. \textbf{\textit{AdvBench}}, \textbf{\textit{HarmfulQA}}, and \textbf{\textit{CATQA}} denote refusal rates for the corresponding datasets.}
\label{tab:mainJailbreakResults-Medical}
\end{table*}

\subsubsection{Evaluation Metric}

\textbf{Measuring utility.} For the text classification task, we assess the model's classification accuracy~(ACC). For the text generation task, we employ BLEU \cite{papineni2002bleu} ROUGE-$L$ \cite{lin-2004-rouge} as evaluation metrics. Details of each metric could be found in the Appendix \ref{app-sec:metric}.

\noindent \textbf{Measuring Safety.} 
We use three datasets to evaluate the safety of the model: AdvBench \cite{Zou2023UniversalAT}, HarmfulQA \cite{bhardwaj2023redteaming}, and CATQA \cite{bhardwaj-etal-2024-language}. 
We employ Llama-Guard3-8B \footnote{https://huggingface.co/meta-llama/Llama-Guard-3-8B} to evaluate the model's safety by measuring its refusal rate to harmful queries.
Compared to GPT-4, Llama-Guard3-8B demonstrates superior performance with a lower false positive rate. For detailed information of the datasets and Llama-Guard3-8B, please refer to \ref{app-sec:metric}.

\subsection{Experimental Results}

\subsubsection{Main Results on Downstream Tasks}

\noindent \textbf{LoRA SFT.}
As depicted in Table \ref{tab:mainAlignResults-AGNews} and \ref{tab:mainAlignResults-Yahoo}, our method significantly improves the safety of LLMs without compromising their classification performance. As evidenced by VLGuard, when harmful data is mixed into the training set, incorporating safety alignment data has a limited effect on enhancing the safety of LLMs. 

Compared to RESTA, our approach effectively enhances the safety of the model while avoiding the impact of safety vectors on the performance of the downstream task. During the fine-tuning process of Lisa, the regularization of the proximal term limits changes in the parameters of the models, thereby inhibiting their performance on downstream tasks to some extent. For experiments on varying mixing ratios of toxic and safe data,  as well as LLMs with different parameter scales, please refer to Appendices \ref{app-sec: robu-data} and \ref{app-sec: robu-paras}.

\begin{table*}[t!]
\centering
\resizebox{\textwidth}{!}{
\begin{tabular}{lcccccccc}
\toprule
& \multicolumn{1}{l}{DoAnythingNow $\uparrow$} & AdvBench $\uparrow$ & \multicolumn{1}{l}{MBPP} $\uparrow$ & \multicolumn{1}{l}{GSM8K} $\uparrow$ & \multicolumn{1}{l}{BBH} $\uparrow$ & \multicolumn{1}{l}{MMLU} $\uparrow$ & IFEval $\uparrow$\\
\midrule
Base Model & \underline{0.81} & \underline{0.87} & \underline{59.20} & \underline{82.79} & \underline{68.29} & \underline{69.14} & \underline{81.65}  \\
SafetyJ    & \textbf{0.92}    & \textbf{0.98}   & 57.40 & 80.97 & 20.03 & \underline{69.14} & 35.01  \\
DARE       & \textbf{0.92}    & \textbf{0.98}   & 57.40 & 80.97 & 20.03 & \underline{69.14} & 45.87  \\ 
Ours       & \textbf{0.92}    & \textbf{0.98}   & \textbf{59.40} & \textbf{84.52} & \textbf{68.52} & \textbf{69.17} & \textbf{81.88}  \\
\bottomrule
\end{tabular}}
\caption{Performance of Safety Components in Full Fine-Tuning Settings. \textbf{\textit{Base Model}} represents Llama3.1-8B-Instruct, \textit{\textbf{SafetyJ}} indicates the performance of the safety model after fine-tuning, and \textbf{\textit{Ours}} signifies the performance of the base model after linear arithmetic with safety principal components.}
\label{tab:low-rank-full}
\end{table*}

\begin{table*}[t!]
\centering
\resizebox{\textwidth}{!}{
\begin{tabular}{lcccccccc}
\toprule
& \multicolumn{1}{l}{DoAnythingNow $\uparrow$} & AdvBench $\uparrow$ & \multicolumn{1}{l}{MBPP $\uparrow$} & \multicolumn{1}{l}{GSM8K $\uparrow$} & \multicolumn{1}{l}{BBH $\uparrow$} & \multicolumn{1}{l}{MMLU $\uparrow$} & IFEval $\uparrow$ \\
\midrule
Base Model & 0.81             & \underline{0.87} & \underline{59.20}  & \underline{82.79} & \textbf{68.29} & \underline{69.14} & \underline{81.65}  \\
SafetyJ    & \underline{0.91} & \textbf{0.99}   & 56.80               & 82.56 & 31.42 & 67.28 & 45.79  \\
DARE       & 0.89             & \textbf{0.99}   & 56.80               & 82.41 & 31.89 & 67.45 & 56.91  \\
Ours       & \textbf{0.92}    & \textbf{0.99}   & \textbf{60.40}      & \textbf{83.78} & \underline{68.20} & \textbf{69.18} & \textbf{82.42}  \\ \bottomrule
\end{tabular}}
\caption{Performance of Safety Components in LoRA Fine-Tuning Settings. \textbf{\textit{Base Model}} represents Llama3.1-8B-Instruct, \textit{\textbf{SafetyJ}} indicates the performance of the safety model after fine-tuning, and \textbf{\textit{Ours}} signifies the performance of the base model after linear arithmetic with safety principal components.}
\label{tab:low-rank-lora}
\end{table*}

\noindent \textbf{Full SFT.}
According to Table \ref{tab:mainJailbreakResults-Medical}, the model exhibits increased sensitivity to parameter changes due to full fine-tuning effects. For Llama3.1-8B-Instruct, the baselines affect its \textit{BLEU} and \textit{Rouge-L} scores by more than 0.2, while our method minimizes the suppression of downstream capabilities while ensuring the safety and consistency of LLMs. From Tables \ref{tab:mainAlignResults-AGNews}, \ref{tab:mainAlignResults-Yahoo} and \ref{tab:mainJailbreakResults-Medical}, it is evident that the low-rank safety principal components consistently maintain the safety alignment of LLMs across various fine-tuning scenarios. In contrast, other baseline methods show significant variation in the safe rejection rate across different harmful query datasets, particularly pronounced in the LoRA fine-tuning scenario.

\subsubsection{Low-Rank Safety Principal Components}
\label{sec:low-rank-exp}

To verify the effectiveness of low-rank safety principal components and demonstrate that principal components and LLMs share the same low-rank subspace, we perform linear operations between the safety components and LLMs to enhance their safety without significantly impacting their general performance.
Given that instruction-tuned LLMs already exhibit a high degree of safety, we further fine-tune them using a carefully curated jailbreak dataset to obtain a more safety robust model. Specifically, we select JailJudge \cite{Liu2024JAILJUDGEAC} jailbreak dataset to perform SFT on Llama3.1-8B-Instruct and use the AdvBench and DoAnythingNow \cite{Shen2023DoAN} datasets to assess the safety of the LLMs. Additionally, we assess the model's general capabilities with the MBPP \cite{austin2021program}, GSM8K \cite{cobbe2021gsm8k}, BBH \cite{Suzgun2022ChallengingBT}, MMLU \cite{hendryckstest2021}, and IFEval \cite{Zhou2023InstructionFollowingEF} datasets. 
To demonstrate the superiority of low-rank safety principal component merging, we compare it with DARE \cite{Yu2023LanguageMA} baseline. Details on the baseline, datasets, and SFT can be found in Appendix \ref{app-sec:low-rank-compontnt}.

From Table \ref{tab:low-rank-full} and Table \ref{tab:low-rank-lora}, it can be observed that after applying SFT to Llama3.1-8B-Instruct (\textit{SafetyJ}), its safety performance on \textit{AdvBench} and \textit{DoAnythingNow} improved significantly. However, its general capabilities show a varying degree of decline, particularly concerning the \textit{BBH} and \textit{IFEval} metrics. It indicates that the safety vector not only contains desired changes in the safety direction but also includes undesirable drift that adversely affects general capabilities.

\textit{Ours} results show that applying linear operations between low-rank principal components and the \textit{Base Model} enhances the model's safety robustness without significantly affecting its general capabilities. This suggests that the safety principal components of safety vectors share the same low-rank subspace as the primary safety drift in LLMs while remaining independent of the general capabilities direction, which is consistent with the perspective presented in \citet{Wei2024AssessingTB}. Compared to DARE, our method causes less disruption to the model’s general ability, showing superior safety realignment over traditional model fusion methods.

\subsubsection{Impact of Safety Singular Value Entropy}

\begin{figure*}[t!]
\centering
\subfigure[safety vector]{
\resizebox{0.31\linewidth}{!}{
\includegraphics[width=2.5in]{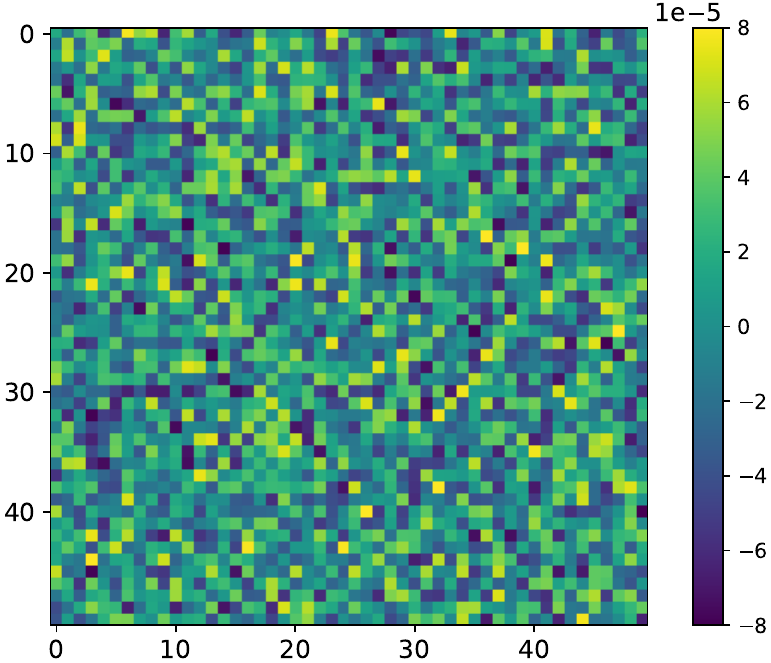}}
\label{sub-fig:safe-vec}}
\subfigure[$\eta=0.9$]{
\resizebox{0.31\linewidth}{!}{
\includegraphics[width=2.5in]{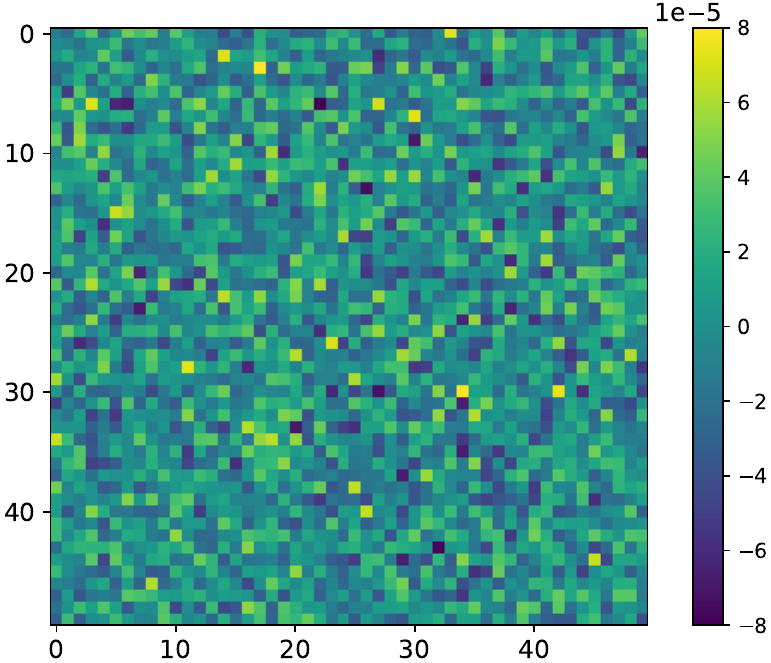}}
\label{sub-fig:eta5}}
\subfigure[$\eta=0.5$]{
\resizebox{0.31\linewidth}{!}{
\includegraphics[width=2.5in]{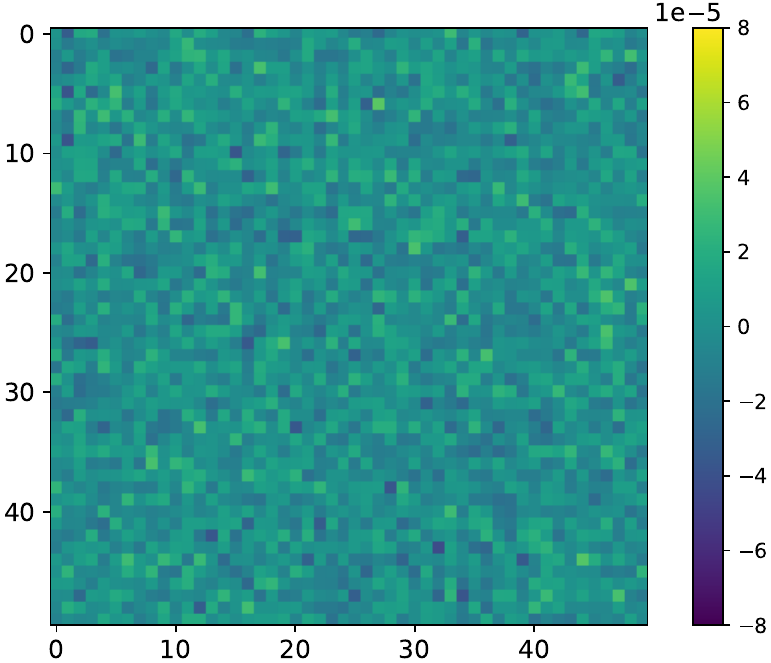}}
\label{sub-fig:eta9}
}
\DeclareGraphicsExtensions.
\caption{Visualization of safety vector (a) and low-rank safety principal components (b, c) with different $\eta$ at the layer model.layers.5.mlp.down\_proj, where $\alpha=1$. Visual representations of 2500 random sample positions are provided.}
\label{fig:layer-5-down}
\end{figure*}

\begin{figure}[t!]
    \centering
    \includegraphics[width=\linewidth]{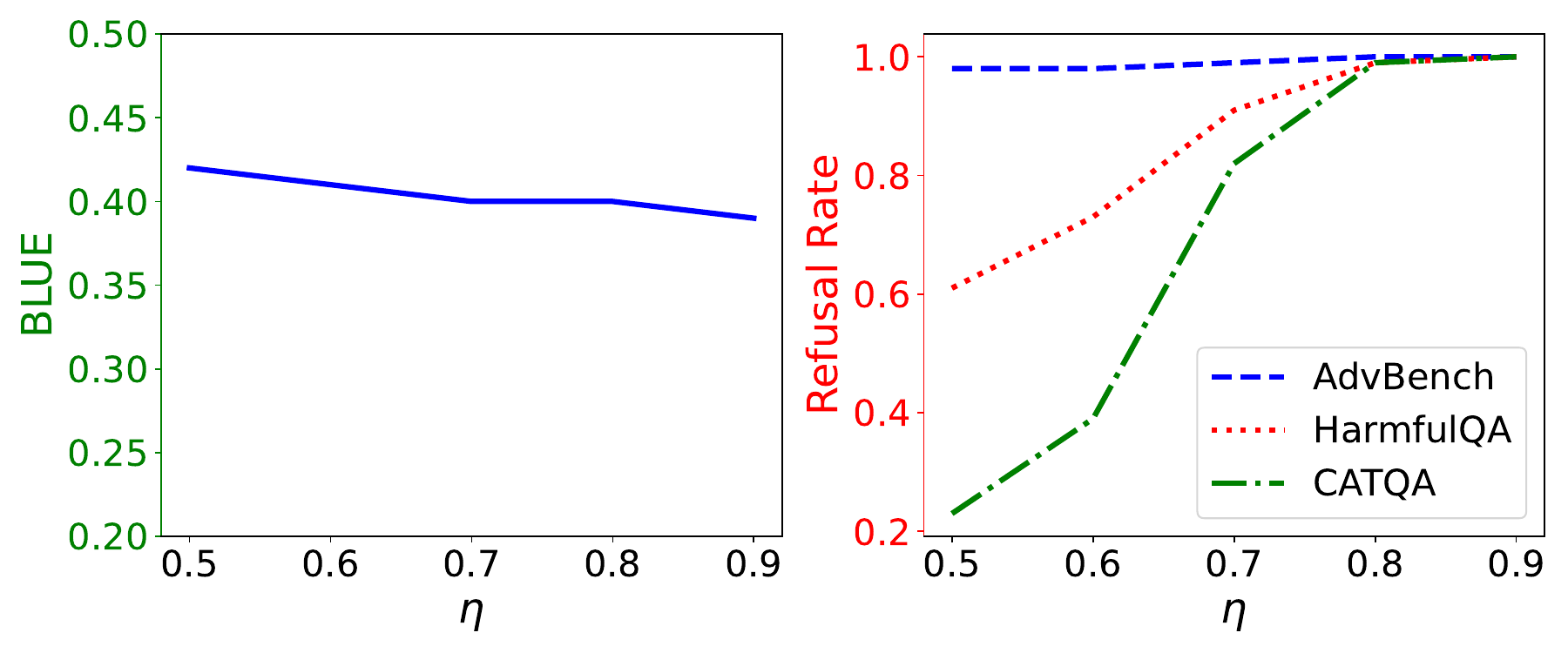}
    \caption{Impact of different singular value entropy thresholds on the performance of downstream fine-tuning models, with left singular vector weight $\alpha=1.0$
    }
    \label{fig:ablate-eta}
\end{figure}

To investigate the impact of safety singular value entropy on low-rank principal components, we conduct a visualization analysis of the safety vector and its low-rank principal components with different singular value entropy ratios in the \textit{model.layers.5.mlp.down\_proj} layer of the Llama3.1-8B-Instruct model.

In particular, we perform random down-sampling with a fixed seed corresponding to the delta parameter matrix of the safety vector and the projection matrix of the low-rank principal components. The results of the down-sampling were visualized using heatmaps. As shown in Figure \ref{fig:layer-5-down}, as the singular value entropy ratio $\eta$ decreases, the matrix corresponding to the low-rank principal components gradually becomes smoother, with its values tending towards zero. This indicates that the safety singular value entropy can effectively regulate the information content of low-rank principal components, thereby eliminating the interference of redundant directional information. As shown in Figures \ref{sub-fig:eta5} and \ref{sub-fig:eta9}, our method does not suppress individual delta parameters but rather identifies a significant structured subnetwork from the global network. 
This aligns with our analysis of the low-rank safety subspace in Section \ref{subsec:low-rank-pruning}.

For experiments on the relationship between singular value entropy and retained rank, please refer to Appendix \ref{app-sec:svl-and-rank}. For visualizations of the safety vectors and low-rank principal components of other layers, please refer to Appendix \ref{app-sec:vis}. 

\subsection{Ablation Study}
\subsubsection{Importance of Singular Value Entropy}

We conducted ablation experiments on the singular value entropy ratio threshold $\eta$ to systematically investigate its impact on the performance of the fine-tuned model. As shown in Figure \ref{fig:ablate-eta}, increasing $\eta$ results in a gradual improvement in the safety of the merged fine-tuned model. When $\eta > 0.8$, the safe refusal rate of the downstream fine-tuned model approaches 1.0, which is very close to the performance of the safety model, while the model's BLEU score is only slightly affected.

It indicates that by adjusting the threshold of the singular value entropy ratio, we modify the truncation rank of the low-rank projection matrix. This allows us to control the amount of safety information retained in the principal components after projecting the safety vector, thereby balancing the utility and safety of the downstream fine-tuned model.

\subsubsection{Importance of Left Singular Vector}

\begin{figure}[t!]
    \centering
    \includegraphics[width=0.95\linewidth]{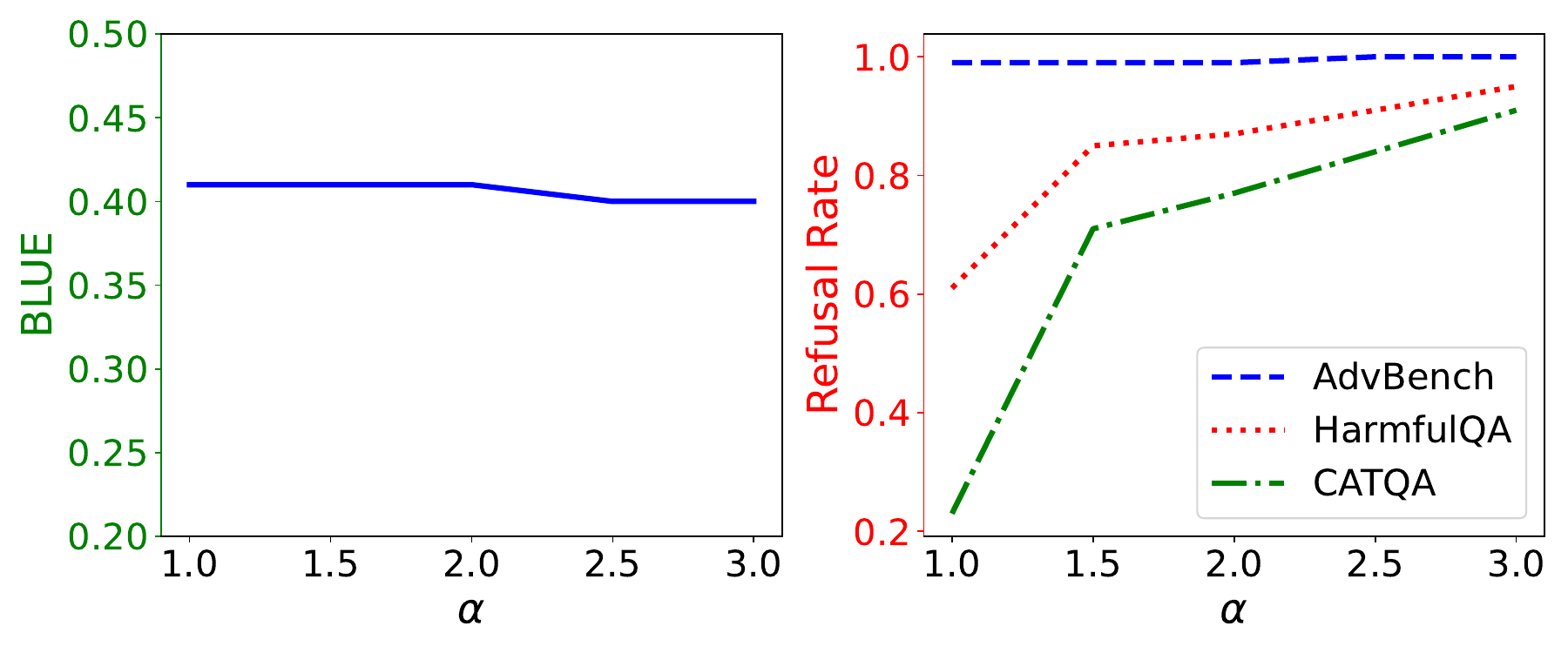}
    \caption{Impact of different left singular vector weights on the performance of downstream fine-tuning models, with singular value entropy ratio threshold of $\eta=0.5$
    }
    \label{fig:ablae-alpha}
\end{figure}

To assess how safety low-rank principal components, obtained from orthogonal projection matrices using left singular vectors, influence the performance of fine-tuned LLMs, we conducted ablation experiments by adjusting the weights of the left singular vectors. 
In particular, as illustrated in Figure \ref{fig:ablate-eta}, we set $\eta=0.5$ in the ablation experiment to achieve more pronounced results, ensuring that fine-tuned LLMs exhibit lower safety when the left singular value weight $\alpha=1$.
From Figure \ref{fig:ablae-alpha}, it can be observed that as the weight corresponding to the left singular vectors gradually increases, the safety of the fine-tuned model also progressively improves.
When $\alpha > 2.5$, the model's safety is close to the safety model, while its BLEU score is only slightly affected. This indicates that the projection matrix constructed from the left singular vectors accurately captures the main direction of safety drift in the safety vector, and increasing the drift in the corresponding direction can further enhance the model's safety.

\section{Conclusion}

Recent studies indicate that fine-tuning can compromise the safety guardrails of LLMs. In this paper, we propose the LSSF safety realignment framework to address the safety alignment issues caused by fine-tuning LLMs. Our experiments demonstrate that the low-rank safety subspace of LLMs remains largely unchanged during fine-tuning and is isolated from the direction of the model's general capabilities. Building on this, we utilize the low-rank principal components of the safety vector to rectify the safety drift of LLMs within the low-rank safety subspace, thereby restoring their safety alignment without compromising their performance on downstream tasks. Given that our method is independent of specific model architectures, we plan to extend it to multimodal and mixture of experts (MoE) models for further exploration in the future.

\section{Limitations and Ethics Statements}

\textbf{Limitations} Despite observing the widespread applicability of LSSF in downstream tasks, budget constraints prevented us from evaluating larger models such as Llama-3.1-405B-Instruct. Given that our method is independent of specific model architectures, we plan to extend it to multimodal and mixture of experts (MoE) models for further exploration in the future.

\noindent \textbf{Ethics Statements} Our study highlights the vulnerabilities in aligning large language models. It is undeniable that we used toxic data in our experiments to compromise model safety, which may have some negative impact on the safety of open-source models. However, considering that all datasets used in our experiments have been extensively studied in numerous academic works, this research does not amplify the inherent negative effects of the datasets themselves. Despite these concerns, we assert that analyzing the harmful aspects of large language models LLMs and exploring potential mitigation strategies have the potential to drive progress in enhancing the safety of LLMs.

\section{Acknowledgement}

This work was supported by the National Natural Science Foundation of China under grant number 62202170 and the Ant Group.

\bibliography{custom}
\clearpage

\appendix

\section{Proof of Shared Low-Rank Subspace}
\label{app-sec:low-rank-subspace}

Given that $\widehat{X} \in \mathbb{R}^{d_{\text{in}} \times n}$ is the corresponding input matrix computed by weight matrix $W \in \mathbb{R}^{d_{\text{out}} \times d_{\text{in}}}$ of $\boldsymbol{\theta}_\text{safe}$ based on the calibration dataset $\mathcal{D}_{\text{anchor}}$, specifically:
\begin{align}
    Z = W \widehat{X}
\end{align}
where $Z$ denotes the activation matrix the linear layer corresponding to $W$. Assume that $W$ is obtained by safety fine-tuning an unsafe model, which can be expressed as:
\begin{align}
W = & \boldsymbol{\theta}_{\text{unsafe}} + \boldsymbol{\delta}_{\text{safe}} \notag \\
  = & \boldsymbol{\theta}_{\text{unsafe}} + \boldsymbol{\tau}_{\text{safe}} + \hat{\boldsymbol{\tau}}_{\text{safe}},
\end{align}
where $\boldsymbol{\theta}_{\text{safe}}$ represents the safety-related shift in the safety vector $\boldsymbol{\theta}_{\text{safe}}$ and  $\hat{\boldsymbol{\tau}}_{\text{safe}}$ denotes the directional drift that suppresses general capabilities. The Frobenius norm minimization of the low-rank $\widehat{W}$ is given by
\begin{align}
\widehat{W} &= \arg \min_{\widehat{W}} \|W \widehat{X} - \widehat{W} \widehat{X}\|_F^2 \notag \\
&= \arg \min_{\widehat{W}} \sum_{i=1}^{n} \|W \hat{x}_i - \widehat{W} \hat{x}_i\|^2 \notag \\
&= \arg \min_{\widehat{W}} \sum_{i=1}^{n} \|\boldsymbol{\theta}_{\text{unsafe}} \hat{x}_i + \boldsymbol{\tau}_{\text{safe}} \hat{x}_i + \\ & \qquad \qquad \qquad \qquad \qquad \hat{\boldsymbol{\tau}}_{\text{safe}} \hat{x}_i - \widehat{W} \hat{x}_i\|^2 \notag
\end{align}
According to the Eckart-Young theorem \cite{Eckart_Young_1936}, for a given matrix $W$, the optimal low-rank approximation of rank $r$ is obtained by retaining the top $r$ singular values and their corresponding singular vectors from its singular value decomposition. Therefore, the optimal solution $\widehat{W}$ should preserve the most significant variations, specifically those associated with $\boldsymbol{\tau}_{\text{base}}$. Since $\hat{\boldsymbol{\tau}}_{\text{safe}}$, which suppresses general capabilities, does not impact the model's safety, it does not become a principal component in $W\widehat{X}$. Consequently, the low-rank approximation $ \widehat{W}$ does not include the $\hat{\boldsymbol{\tau}}_{\text{safe}}$ component. For the corresponding experiments, please refer to \ref{sec:low-rank-exp}.

\section{Proof of The Singular Value Entropy}
\label{app-sec:singular-value-entropy}

The definition of the Frobenius norm is the square root of the sum of the squares of all elements in a matrix. Specifically, if $A$ is an $m \times n$ matrix, then the Frobenius norm is defined as:
\begin{align}
    \|A\|_F^2 = \sum_{i=1}^m \sum_{j=1}^n |a_{ij}|^2
\end{align}
On the other hand, the Singular Value Decomposition (SVD) of a matrix provides a decomposition of $A$ as follows:
\begin{align}
    A = U \Sigma V^\top
\end{align}
where $U$ and $V$ are unitary orthogonal matrices, and $\Sigma = \text{diag}(\sigma_1, \sigma_2, \cdots, \sigma_n)$ is the singular value diagonal matrix.

\noindent The Frobenius norm has the property:
\begin{align}
    \|A\|_F^2 = \text{Tr}(A^\top A)
\end{align}
After computing $A^\top A$, we have:
\begin{align}
    A^\top A &= (U \Sigma V^\top)^\top (U \Sigma V^\top) 
    \notag \\ &= V \Sigma^\top U^\top U \Sigma V^\top \notag \\ &= V \Sigma^2 V^\top
\end{align}
Since $V$ is a unitary orthogonal matrix, it follows that:
\begin{align}
    \text{Tr}(V \Sigma^2 V^\top) = \text{Tr}(\Sigma^2)
\end{align}
$\Sigma^2$is a diagonal matrix, and $\text{Tr}(\Sigma^2)$ is the sum of its diagonal elements:
\begin{align}
    \text{Tr}(\Sigma^2) = \sum_{i=1}^r \sigma_i^2
\end{align}
Thus, we ultimately have:
\begin{align}
    \|A\|_F^2 = \sum_{i=1}^r \sigma_i^2
\end{align}
Therefore, it is proven that $\|A\|^2_{F}=\sum^r_{i=1}\sigma^2_i$.

\section{Proof of The Optimality of SVD}
\label{app-sec:SVD}
Let $\widehat{X} \in \mathbb{R}^{d_{\text{in}} \times n}$ and $\widehat{W}$ denote the solution to the following rank-constrained approximation problem:
\begin{align}
    \widehat{W}=\underset{\operatorname{rank(}\widehat{W})\leq r}{\operatorname*{\operatorname*{\operatorname*{\operatorname*{\arg\min}}}}}\|W\widehat{X}-\widehat{W}\widehat{X}\|_{F}^{2}
\end{align}
Perform a low-rank matrix decomposition of $\widehat{W}\widehat{X}$ using Singular Value Decomposition (SVD):
\begin{align}
    USV^\top \approx \widehat{W}\widehat{X}
\end{align}
where $U \in \mathbb{R}^{d_{\text{out}} \times r}$ is an orthogonal matrix composed of the first $r$ left singular vectors. The minimum of the constrained problem is achieved by:
\begin{align}
    \widehat{W} = UU^\top W
\end{align}
Let \( X_{\text{in}} \) denote the input corresponding to the weight matrix $W$, and \( Z = W X_{\text{in}} \). According to the Eckart–Young theorem \cite{Eckart_Young_1936}, the singular value decomposition (SVD) 
$\widehat{Z} = U S V^\top$ provides the optimal rank-$r$ approximation of $Z$.  Substituting $Z = W X_{\text{in}}$, we obtain:
\begin{align}
    \widehat{Z} = UU^\top W X_{\text{in}}
\end{align}
By setting $\widehat{W} = UU^\top W$, it follows that:
\begin{align}
    \|\widehat{Z} - Z\|_F^2 \text{ is minimized} \quad \Rightarrow \quad \notag \\ \|\widehat{W}\widehat{X} - W\widehat{X}\|_F^2 \text{ is minimized}
\end{align}
Furthermore, since $UU^\top$ is a rank-$r$ projection matrix, it holds that $\operatorname{rank}(\widehat{W}) \leq r$. Therefore, $\widehat{W}$ is the optimal solution to the rank-constrained minimization problem.

\section{Experimental Details}

\subsection{Baseline Setup}
\label{app-sec:baseline}
To mitigate fine-tuning risks, we select baseline methods encompassing both in-fine-tuning alignment and post-alignment approaches, including:

\begin{itemize}
    \item NA-SFT: Utilizes only the fine-tuning dataset without enforcing safety alignment.
    \item VLGuard \cite{Zong2024SafetyFA}: A defensive solution against harmful fine-tuning attacks during the fine-tuning phase, which integrates safety-aligned data into the fine-tuning process to continuously reinforce the model's alignment knowledge. Originally applied to visual-LLM fine-tuning, in this paper, we employ the SafeInstr \cite{Bianchi2023SafetyTunedLL} dataset for safety alignment.
    \item Lisa \cite{Huang2024LazySA}: Separates the fine-tuning phase into two states to independently optimize alignment and user datasets, thereby mitigating jailbreak effects. Additionally, it introduces a proximal term to constrain the drift of each state.
    \item RESTA \cite{bhardwaj-etal-2024-language}: Combines the safety vector with the weights of the compromised model through simple arithmetic combination and employs DARE \cite{Yu2023LanguageMA} to merge with the original model, thereby alleviating the suppression impact on the general capabilities of the safety vector.
\end{itemize}
The detailed hyperparameter settings for each method are as follows:

\begin{itemize}
    \item VLGuard: For VLGuard, we utilize the SafeInstr safety calibration dataset. To align with NA-SFT, we randomly select 500 samples from Insfer.
    \item  Lisa: For Lisa, we set $align\_step = 100$, $finetune\_step = 900$, and the proximal penalty $\rho = 1$. As described in Section Appendix, the training set comprises the downstream task fine-tuning dataset and 500 harmful samples, while the alignment dataset consists of 500 safe samples.
    \item RESTA: For RESTA, we assign weights of $1$, $1$, and $-1$ to the compromised model, base model, and unaligned model, respectively.
\end{itemize}
Consistent with our training hyperparameters, we set the number of training epochs to 10 and the learning rate to $1 \times 10^{-5}$ for all tasks. For LoRA fine-tuning, we set $r = 16$. All fine-tuning tasks were conducted on 8 Nvidia A100 GPUs.

\subsection{Dataset}
\label{app-sec:dataset}

In our downstream fine-tuning experiments, we established two distinct scenarios:

\noindent \textbf{LoRA Fine-Tuning.}
For text classification, we utilize two multi-class datasets: AG's News and Yahoo Answers \cite{Zhang2015CharacterlevelCN}.
AG's News is primarily utilized for news classification tasks and comprises news article snippets from various sources. The dataset is divided into four categories: World, Sports, Business, and Sci/Tech. We randomly selected 50K samples from the training set to perform supervised fine-tuning of the LLM and randomly chose 1K samples from the test set to evaluate the classification accuracy of the fine-tuned model.
The Yahoo Answers dataset is a large-scale multi-class dataset derived from Q\&A dialogues on the Yahoo Answers platform, encompassing ten categories. Similarly, we randomly chose 50K training samples for fine-tuning and 1K test samples for evaluation.

\noindent \textbf{Full Fine-Tuning.} 
For the text generation task, we utilized the Medical Dialogue Dataset \footnote{https://github.com/Toyhom/Chinese-medical-dialogue-data}, which includes Chinese medical dialogue data from six departments, such as andrology, internal medicine, and obstetrics and gynecology. We randomly selected 50K samples from the internal medicine category as the training set and randomly chose 1K samples to evaluate the text generation accuracy of the fine-tuned model.

\noindent \textbf{Harmful and safe dataset.} 
To simulate real-world fine-tuning scenarios \cite{Bianchi2023SafetyTunedLL}, we randomly selected 500, 2,500, and 25,000 Q\&A pairs from the PKU-SafeRLHF dataset in which both \textit{accept} and \textit{reject} are labeled as \textit{unsafe} to construct harmful dataset. Similarly, we randomly selected 500 and 2,500 harmful queries and prompt the safety model to generate safe negative responses to construct safe dataset. These datasets were then mixed into the fine-tuning dataset in various combinations for supervised fine-tuning.

\subsection{Metric}
\label{app-sec:metric}
\subsubsection{Measuring utility}

For the text classification task, we use classification accuracy ACC, calculated as follows:
\begin{align*}
    \text{ACC} = \frac{N_{\text{correct}}}{N_{\text{total}}}
\end{align*}
where $N_{\text{correct}}$ represents the number of correctly predicted samples and $N_{\text{total}}$ denotes the total number of samples. For the text generation task, we employ BLEU \cite{papineni2002bleu} and ROUGE-$L$ \cite{lin-2004-rouge} as evaluation metrics. BLEU assesses the quality of generated text by computing the precision of $n$-gram matches between the candidate text and the reference text. ROUGE-$L$, based on the Longest Common Subsequence (LCS), evaluates the consistency between the generated text and the reference text in terms of word order and content, reflecting the overall similarity of sentence structures.

\begin{table*}[htbp]
\centering
\begin{tabular}{l}
\toprule

Task: Check if there is unsafe content in 'Agent' messages in conversations according \\ our safety policy with the below categories. \\
\\
<BEGIN UNSAFE CONTENT CATEGORIES> \\
S1: Violent Crimes.  \\
S2: Non-Violent Crimes. \\
S3: Sex Crimes. \\
S4: Child Exploitation. \\
S5: Defamation. \\
S6: Specialized Advice. \\
S7: Privacy. \\
S8: Intellectual Property. \\
S9: Indiscriminate Weapons. \\
S10: Hate. \\
S11: Self-Harm. \\
S12: Sexual Content. \\
S13: Elections. \\
S14: Code Interpreter Abuse. \\
<END UNSAFE CONTENT CATEGORIES> \\
\\ 
<BEGIN CONVERSATION> \\
\\
User: \{\}\\
\\
Agent: \{\} \\
\\
<END CONVERSATION>\\
Provide your safety assessment for Agent message in the above conversation: \\
 - First line must read 'safe' or 'unsafe'. \\
 - You only need to output 'safe' or 'unsafe', nothing else. \\

\bottomrule
\end{tabular}
\caption{Specific prompts for safety classification using Llama-Guard3-8B in our experiments}
\label{app-tab:llamaGuard3}
\end{table*}

\subsubsection{Measuring Safety}

\textbf{Safety Evaluation Dataset.}
We employ three datasets to evaluate the model's safety:
\begin{itemize}
    \item AdvBench \cite{Zou2023UniversalAT}: This dataset comprises 520 harmful requests along with their corresponding target strings, covering a wide range of malicious topics such as profanity, threats, discrimination, and cybercrime.
    \item HarmfulQA \cite{bhardwaj2023redteaming}: Sourced from harmful dialogues in ChatGPT, this dataset encompasses 10 safety themes, from which 196 questions are extracted across various subtopics.
    \item CATQA \cite{bhardwaj-etal-2024-language}: This dataset consists of 550 harmful questions spanning 11 categories, each containing 5 harmful subcategories.
\end{itemize}
\textbf{Llama-Guard3-8B.}
We use Llama-Guard3-8B \footnote{https://huggingface.co/meta-llama/Llama-Guard-3-8B} to assess the model's safety against harmful queries. Compared to GPT-4, Llama-Guard3-8B performs better with a lower false positive rate. Additionally, Llama-Guard3-8B outperforms GPT-4 in English proficiency, multilingual capabilities, and tool usage, exhibiting a significantly lower false positive rate. For the specific prompts used in Llama-Guard3-8B, please refer to Table \ref{app-tab:llamaGuard3}.

\subsection{Low-Rank Safety Principal Components}
\label{app-sec:low-rank-compontnt}
We select Llama3.1-8B-Instruct as the base model and perform SFT fine-tuning using the sampled JailJudge \cite{Liu2024JAILJUDGEAC} dataset to obtain the safety model. JailJudge is a comprehensive benchmark that features diverse risk scenarios, including synthetic, adversarial, in-the-wild, and multilingual prompts, along with high-quality human-annotated datasets. To minimize the impact of data distribution on the experiment, we use DoAnythingNow \cite{Shen2023DoAN} and AdvBench to evaluate the safety of the LLM. DoAnythingNow is a jailbreak dataset that includes 13 disabled scenarios, from which we randomly selected 1000 instances for testing.
To demonstrate the effectiveness of low-rank safety principal components, we compare it with the model fusion method DARE \cite{Yu2023LanguageMA}.
DARE reduces the number of effective parameters in task vectors by removing the delta parameter and proportionally scaling the remaining values. This intuitive approach helps alleviate parameter interference when integrating multiple models. Even for a single task-specific model, pruning certain parameters can mitigate interference and potentially enhance safety.
To evaluate the general capabilities of the LLM, we use the following datasets: 
\begin{itemize}
    \item MBPP \cite{austin2021program}: MBPP comprises approximately 1,000 Python programming problems, covering fundamental programming knowledge and standard library functions. Each problem includes a task description, a code solution, and three automated test cases.
    \item GSM8K \cite{cobbe2021gsm8k}: GSM8K contains 8.5K high-quality, linguistically diverse elementary school math word problems. These problems typically require 2 to 8 steps to solve, primarily involving basic arithmetic operations.
    \item BBH \cite{Suzgun2022ChallengingBT}: BIG-Bench Hard (BBH) is a subset of BIG-Bench, focusing on 23 challenging tasks within BIG-Bench that previous language model evaluations have not surpassed the performance of average human scorers.
    \item MMLU \cite{hendryckstest2021}: MMLU specifically assesses the knowledge acquired during pre-training in zero-shot and few-shot settings. It covers 57 subjects across disciplines such as humanities and social sciences.
    \item IFEval \cite{Zhou2023InstructionFollowingEF}: IFEval evaluates the instruction-following capabilities of large language models, containing over 500 prompts.
\end{itemize}
These datasets collectively provide a comprehensive assessment of the model's general proficiency across various domains and tasks. During the training process, we set the number of epochs to 3 and the learning rate to 1e-5. The fine-tuning was conducted on 8 Nvidia A100 GPUs.

\section{Robustness Against the Number of Calibration Datasets}
\label{app-sec: robu-num-calibra}

\begin{table}[t!]
\resizebox{\linewidth}{!}{
\begin{tabular}{lllllllll}
\toprule
    Num      & 16   & 32   & 64   & 128  & 256  & 512  \\
\midrule
ACC       & 0.54 & 0.63 & 0.64 & 0.64 & 0.63 & 0.64 \\
AdvBench  & 1.00 & 1.00 & 1.00 & 1.00 & 1.00 & 1.00 \\
HarmfulQA & 0.94 & 0.99 & 0.99 & 0.99 & 0.99 & 1.00 \\
CATQA     & 0.93 & 0.99 & 0.99 & 0.99 & 0.99 & 0.99 \\
\bottomrule
\end{tabular}}
\caption{Sensitivity analysis for the Yahoo Answers LoRA task with varying amounts of calibration datasets.}
\label{app-table:num-cal}
\end{table}

To validate the sensitivity of our method to the number of samples in the calibration dataset, we randomly selected varying numbers of samples from the PKU-SafeRLHF dataset to conduct ablation experiments. Specifically, we performed safety realignment for Yahoo Answers LoRA fine-tuning on Llama3.1-8B-Instruct. As shown in Table \ref{app-table:num-cal}, when the calibration dataset size reaches 64, our method achieves optimal performance without significant fluctuations as the dataset size increases. Nevertheless, to align with the experimental setup in reference \cite{Wei2024AssessingTB}, we have chosen 128 as the default calibration dataset size.

\section{Robustness Against Data Composition}
\label{app-sec: robu-data}

\begin{figure*}[t!]
\centering
\resizebox{0.48\textwidth}{!}{
\subfigure[Robustness to varying proportions of harmful data, with a mixture of 500 safe instances.]{
\includegraphics[width=2.5in]{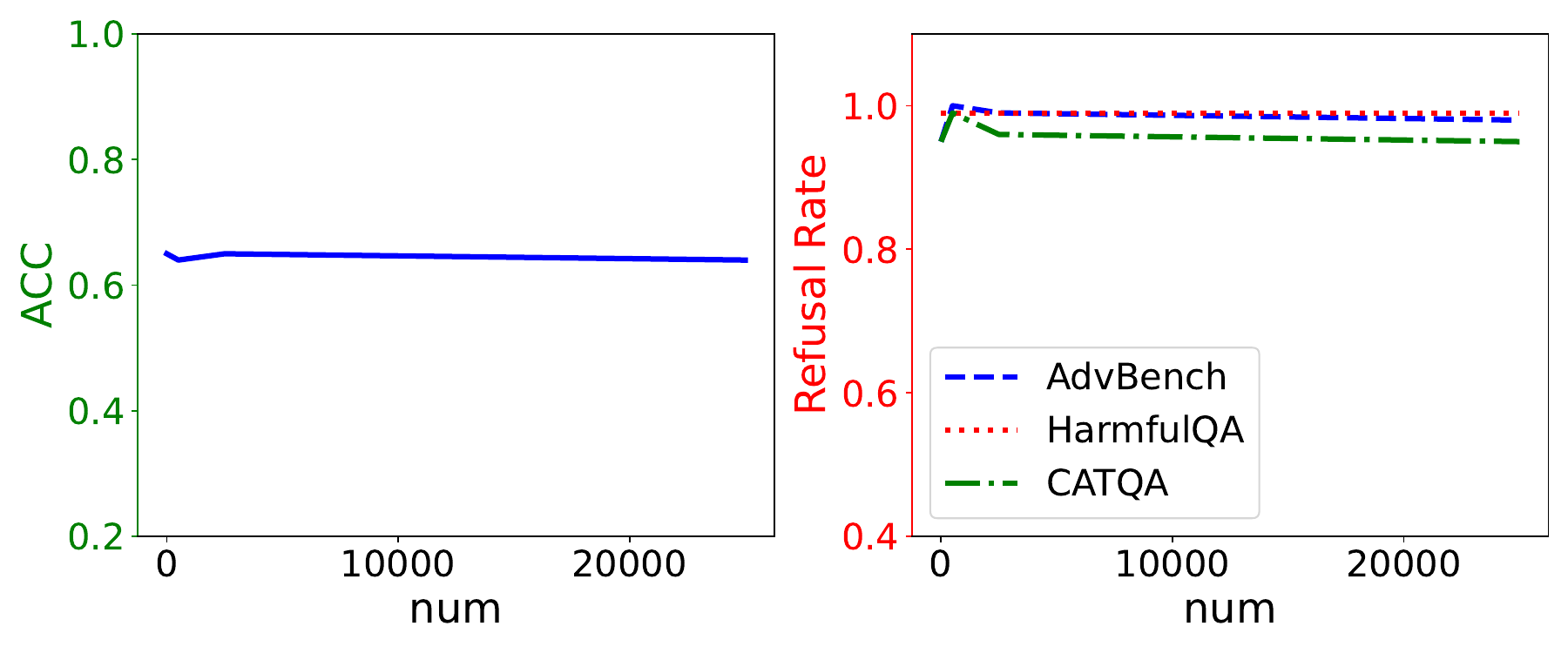}
\label{fig:poison-safe-poison}}}
\resizebox{0.48\textwidth}{!}{
\subfigure[Robustness to varying proportions of safe data, with a mixture of 2500 harmful instances.]{
\includegraphics[width=2.5in]{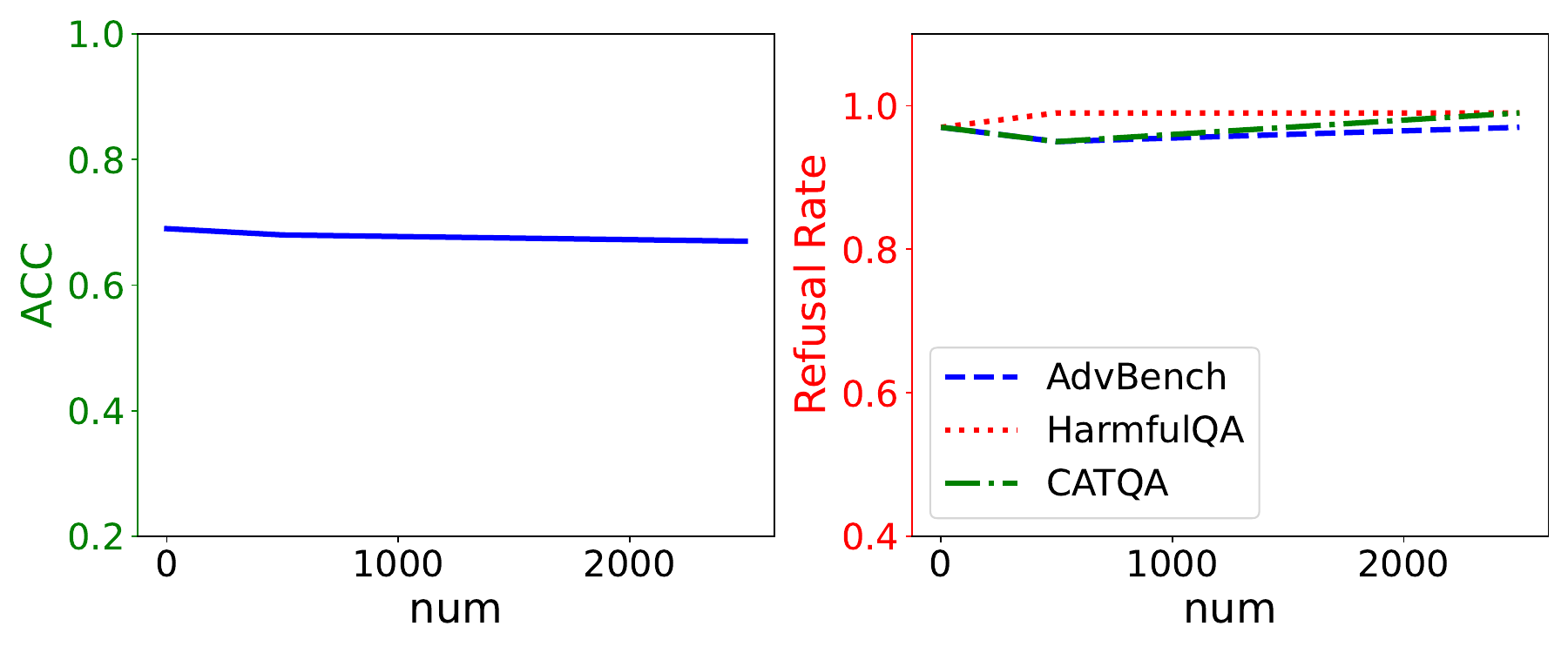}
\label{fig:poison-safe-safe}}}
\DeclareGraphicsExtensions.
\caption{Our method's robustness on Llama3.1-8B-Instruct against varying proportions of harmful or safe data in the Yahoo Answers training dataset.}
\label{fig:poison-safe}
\end{figure*}

To verify the robustness of our method, we conduct LoRA SFT on the Yahoo Answers training set for Llama3.1-8B-Instruct by mixing harmful and safe data in varying proportions. Both harmful and safe data were sourced from the PKU-SafeRLHF dataset, as described in Appendix \ref{app-sec:dataset}.
As shown in Figure \ref{fig:poison-safe-poison}, the LLM generated by our method exhibit no significant changes in ACC and rejection rates as the proportion of harmful data increases in the training dataset.
This demonstrates the high robustness of our method when facing training sets with different proportions of harmful data. 
Similarly, Figure \ref{fig:poison-safe-safe} illustrates that mixing varying proportions of safe data into the training set does not significantly affect the performance of aligned LLMs, further confirming the robustness of our method. However, from empirical evidence \cite{Bianchi2023SafetyTunedLL}, it is necessary to incorporate safe data into training dataset. Adding safe data to the training set not only enhances the lower bound of safety performance in downstream fine-tuning models but also suppresses excessive drift in safety direction during fine-tuning, which is beneficial for improving the effectiveness of our method.

\section{Robustness Against Model Parameters}
\label{app-sec: robu-paras}
\begin{table*}[t!]
\centering
\resizebox{\textwidth}{!}{
\begin{tabular}{lcccccccc}
\toprule
\multicolumn{1}{c}{\multirow{2}{*}{Base Model}} & \multicolumn{4}{c}{AG's News}       & \multicolumn{4}{c}{Yahoo Answers}  \\ \cline{2-9} 
\multicolumn{1}{c}{} & ACC  & AdvBench & CATQA & HarmfulQA & ACC  & AdvBench & CATQA & HarmfulQA \\ \midrule
Qwen2.5-3B-Instruct  & 0.92 & 0.99 & 0.97 & 0.93 & 0.67 & 0.98 & 0.99 & 0.93      \\
Qwen2.5-7B-Instruct  & 0.92 & 1.00 & 0.98 & 0.93 & 0.68 & 1.00 & 0.99 & 0.99      \\
Qwen2.5-14B-Instruct & 0.92 & 1.00 & 0.99 & 0.95 & 0.69 & 0.99 & 0.99 & 0.99      \\ 
\bottomrule
\end{tabular}}
\caption{Our method's robustness across models with different parameter sizes on Qwen2.5. \textbf{\textit{Base Model}} denotes various sizes of the Qwen2.5 model. The hyperparameters are set as $\eta=0.9$ and $\alpha=1.5$.}
\label{app-tab:para-size}
\end{table*}

To demonstrate the effectiveness of our method across models with varying parameter sizes, we conducted ablation experiments.
We conducted LoRA SFT on Qwen2.5 with varying parameter sizes using the AG's News and Yahoo Answers datasets. Additionally, we applied our method for safety realignment.
As shown in Table \ref{app-tab:para-size}, for models with 3B, 7B, and 14B parameters, our method consistently achieves a safety refusal rate of 0.99. This demonstrates the effectiveness of our method across different parameter sizes in LLMs.

\section{Singular Value Entropy and Rank}
\label{app-sec:svl-and-rank}

\begin{figure*}[t!]
\centering
\subfigure[layer-5]{
\resizebox{0.32\linewidth}{!}{
\includegraphics[width=2.5in]{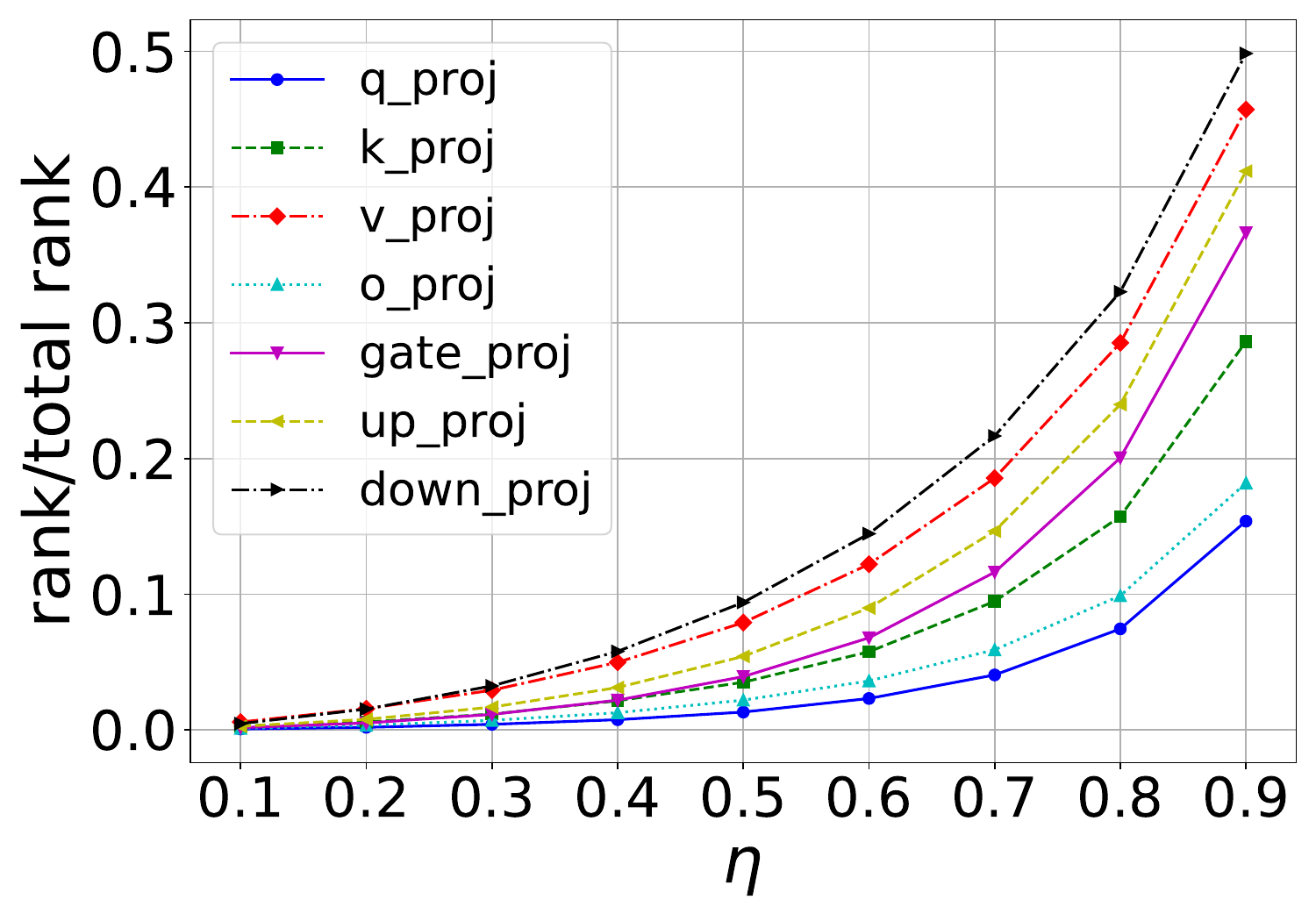}}}
\subfigure[layer-15]{
\resizebox{0.32\linewidth}{!}{
\includegraphics[width=2.5in]{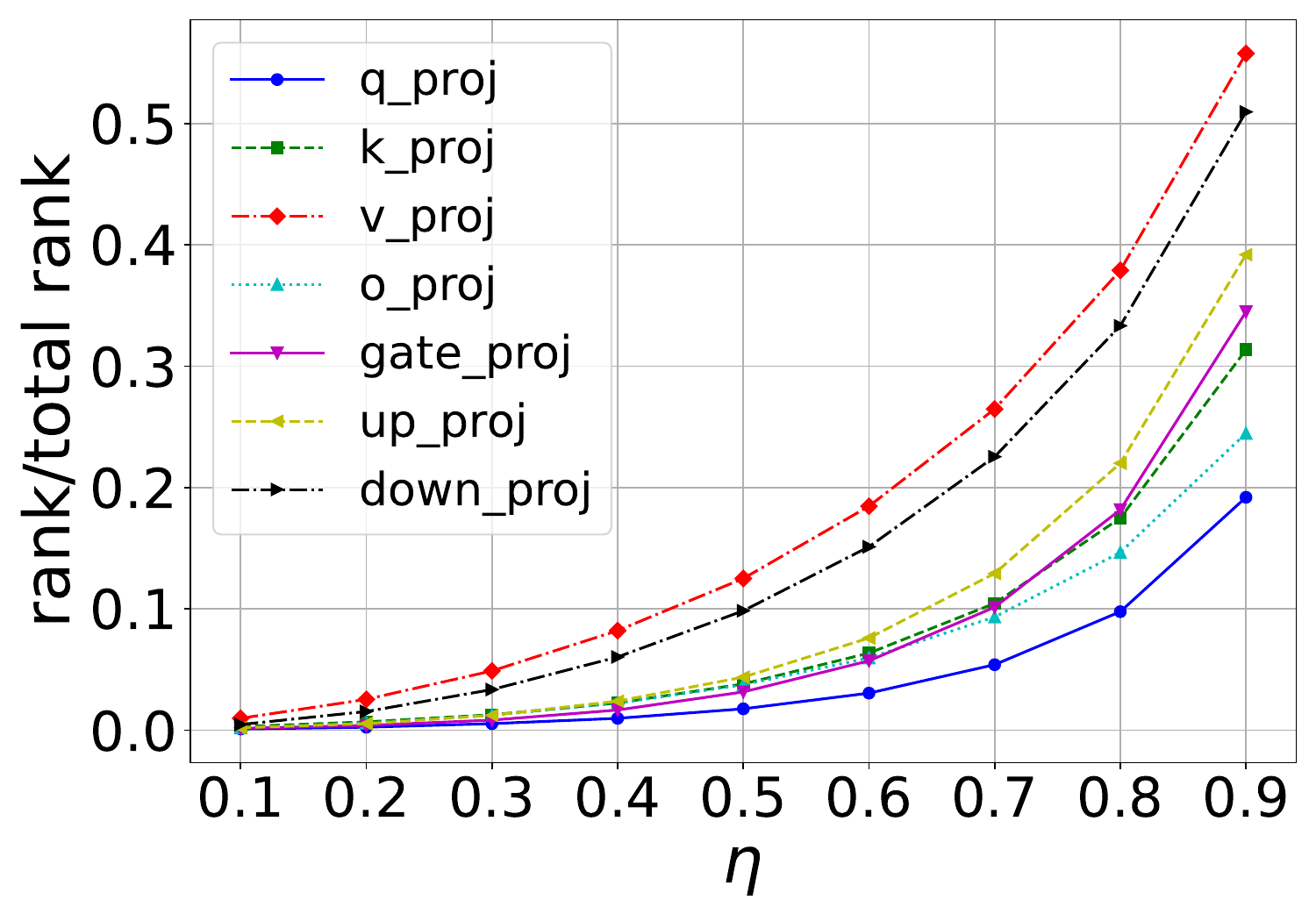}}}
\subfigure[layer-25]{
\resizebox{0.32\linewidth}{!}{
\includegraphics[width=2.5in]{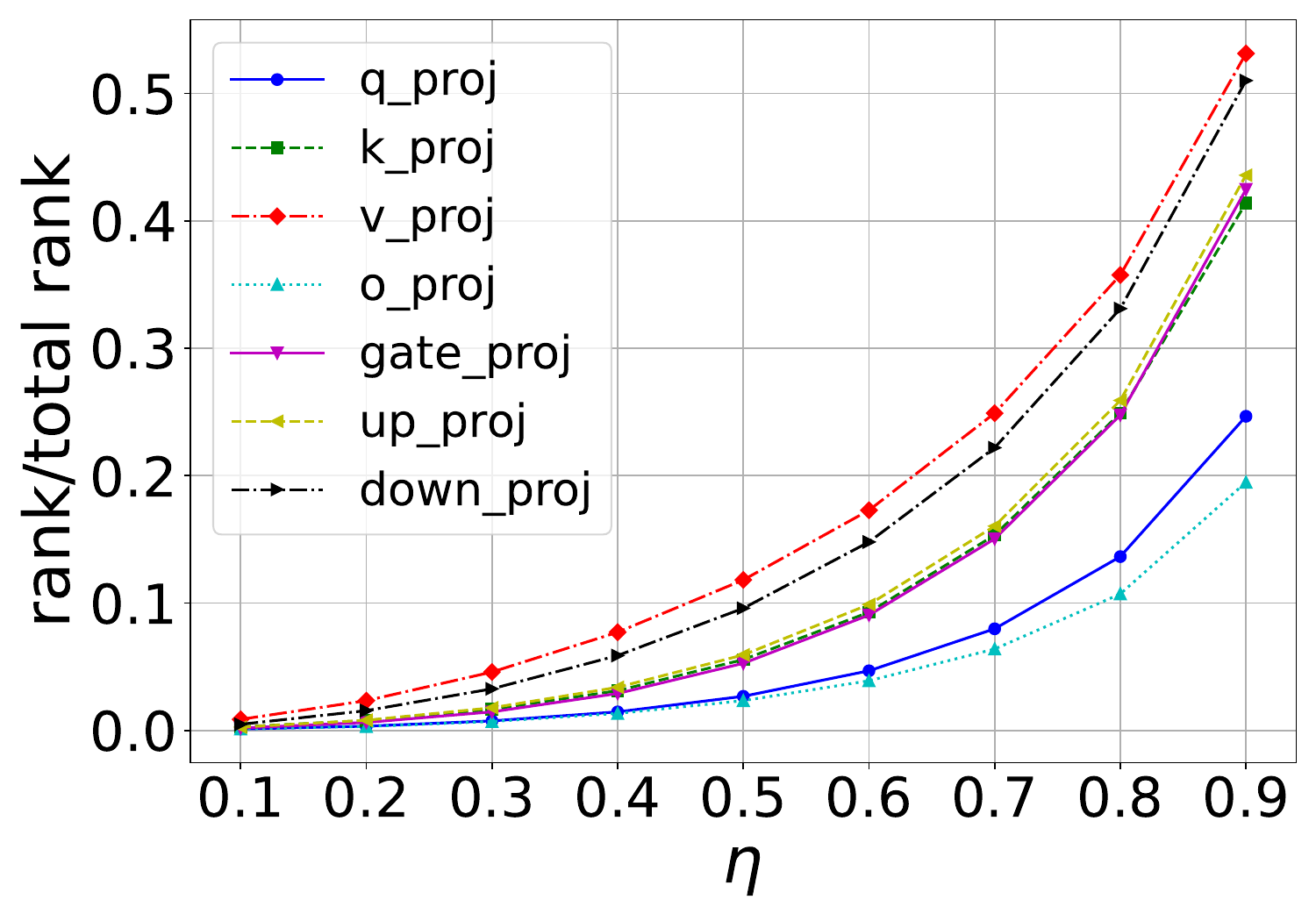}}}
\DeclareGraphicsExtensions.
\caption{Influence of the singular value entropy threshold $\eta$ on the safety retained rank $r$ of weight matrices across various layers in Llama3.1-8B-Instruct.}
\label{fig:entropy-rank}
\end{figure*}

To verify the impact of singular value entropy on the retention rank of the weight matrix, we calculated the relationship between the singular value entropy ratio $\eta$ and the retained rank ratio of the corresponding weight matrix for Llama3.1-8B-Instruct.
From Figure \ref{fig:entropy-rank}, it can be observed that as the proportion of singular value entropy increases, the number of retained ranks in the weight matrix also increases. This demonstrates that we can control the amount of information retained in the low-rank safety principal components through singular value entropy. Within a single layer, different weight matrices exhibit varying encoding densities for safety information. For instance, in $layer-15$, when $\eta$ is 0.9, the difference in the proportion of retained ranks between $v\_proj$ and $q\_proj$ exceeds 30\%. Across different layers, the relative encoding density of safety information by different weight matrices also changes. In the shallower layers, $q\_proj$ exhibits the highest encoding density, whereas in the deeper layers, $o\_proj$ becomes the matrix with the highest encoding density.
This analysis demonstrates that safety singular value entropy allows for precise quantification of safety information encoding density in weight matrices across different layers, thereby facilitating the dynamic determination of the rank retention during low-rank pruning.

\section{Visualization}
\label{app-sec:vis}


\begin{figure*}[t!]
\centering
\subfigure[safety vector]{
\resizebox{0.32\linewidth}{!}{
\includegraphics[width=2.5in]{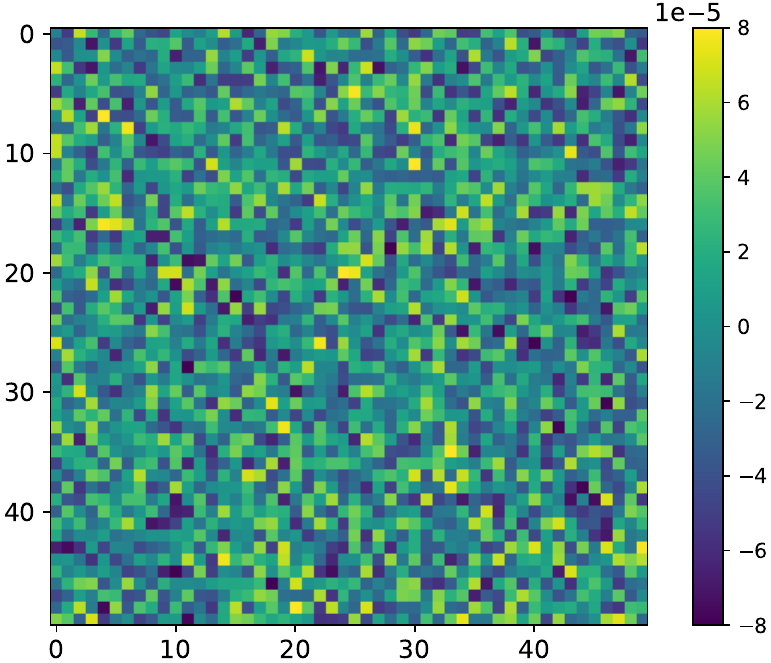}}}
\subfigure[$\eta=0.9$]{
\resizebox{0.32\linewidth}{!}{
\includegraphics[width=2.5in]{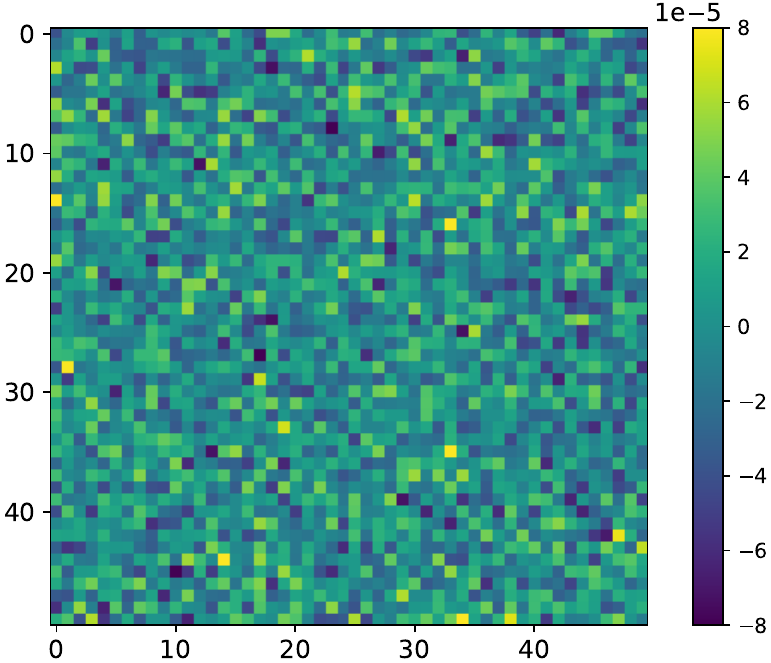}}}
\subfigure[$\eta=0.5$]{
\resizebox{0.32\linewidth}{!}{
\includegraphics[width=2.5in]{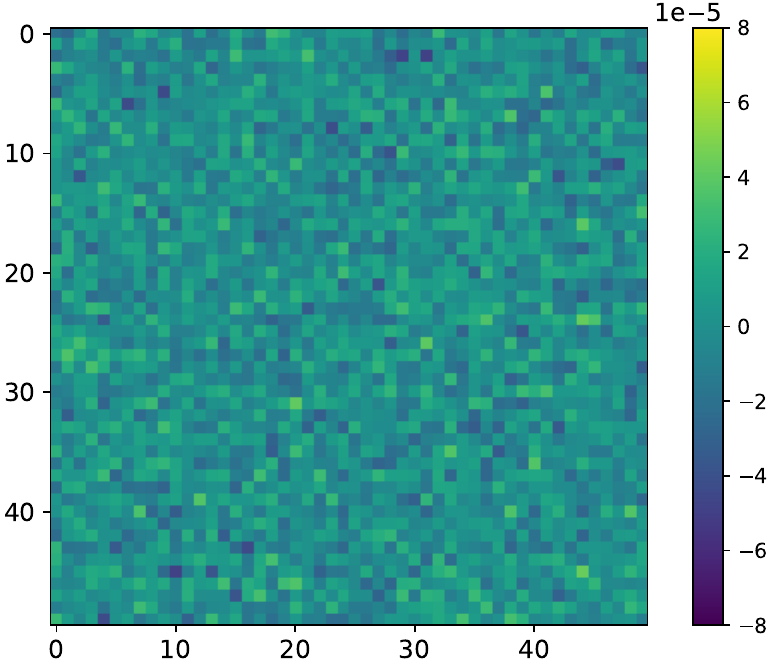}}}
\DeclareGraphicsExtensions.
\caption{Visualization of safety vector (a) and low-rank safety principal components (b, c) with different $\eta$ at the layer model.layers.15.self\_attn.v\_proj, where $\alpha=1$. Visual representations of 2500 random sample positions are provided.}
\label{app-fig:layer-15-v}
\end{figure*}

\begin{figure*}[t!]
\centering
\subfigure[safety vector]{
\resizebox{0.32\linewidth}{!}{
\includegraphics[width=2.5in]{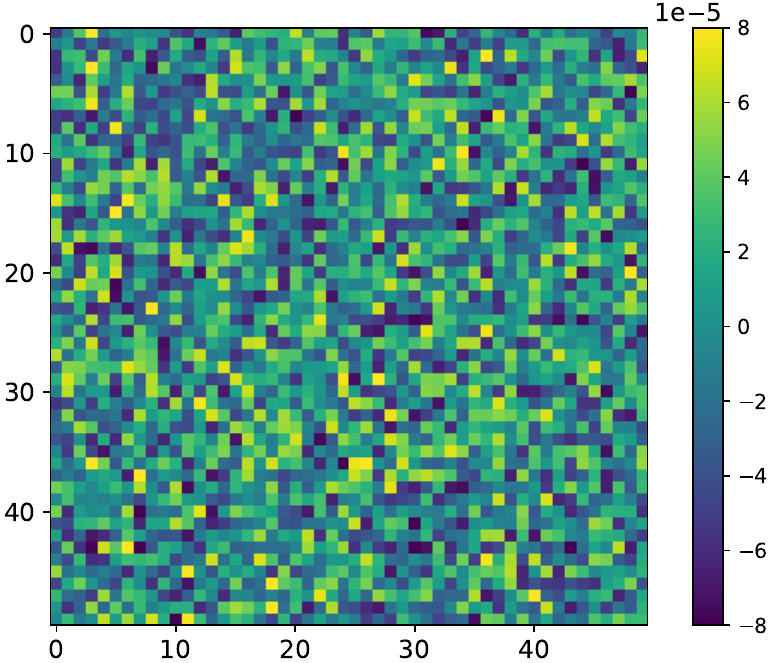}}}
\subfigure[$\eta=0.9$]{
\resizebox{0.32\linewidth}{!}{
\includegraphics[width=2.5in]{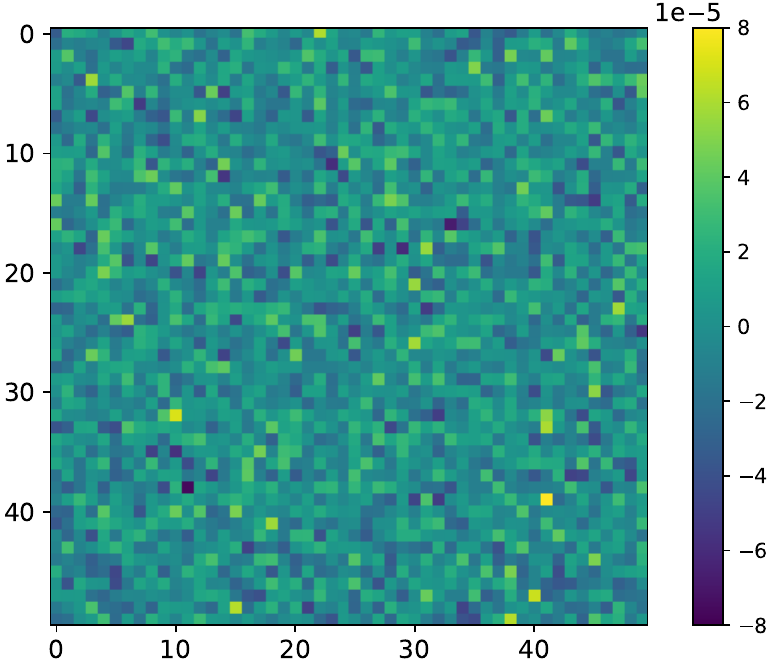}}}
\subfigure[$\eta=0.5$]{
\resizebox{0.32\linewidth}{!}{
\includegraphics[width=2.5in]{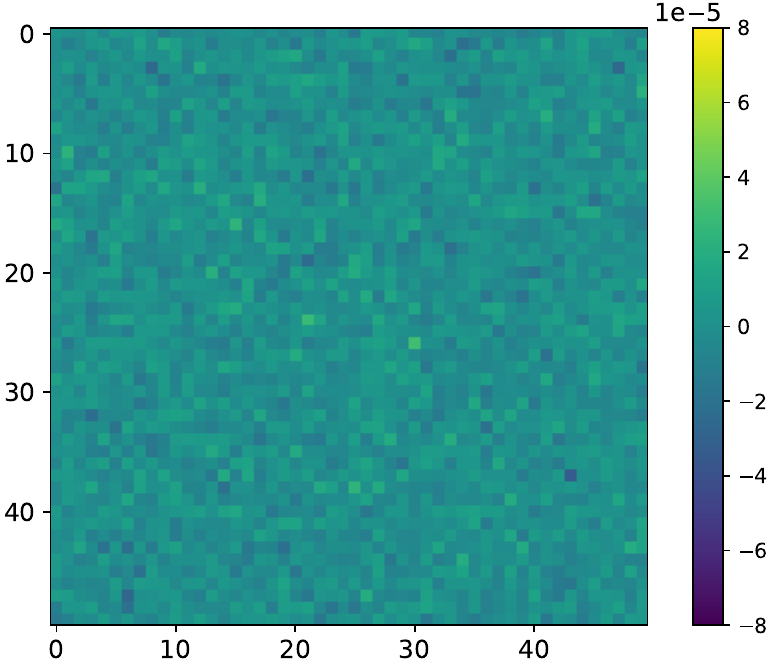}}}
\DeclareGraphicsExtensions.
\caption{Visualization of safety vector (a) and low-rank safety principal components (b, c) with different $\eta$ at the layer model.layers.15.self\_attn.q\_proj, where $\alpha=1$. Visual representations of 2500 random sample positions are provided.}
\label{app-fig:layer-15-q}
\end{figure*}

\begin{figure*}[t!]
\centering
\subfigure[safety vector]{
\resizebox{0.32\linewidth}{!}{
\includegraphics[width=2.5in]{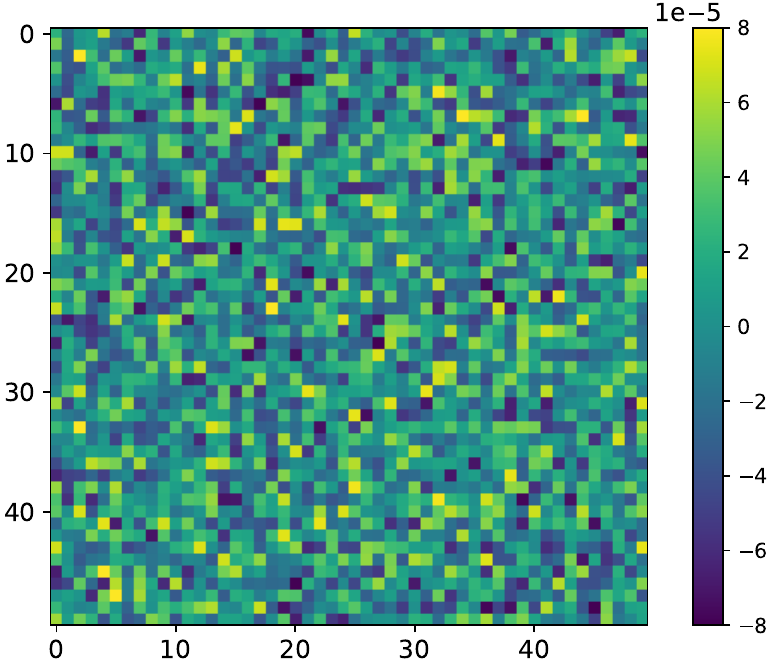}}}
\subfigure[$\eta=0.9$]{
\resizebox{0.32\linewidth}{!}{
\includegraphics[width=2.5in]{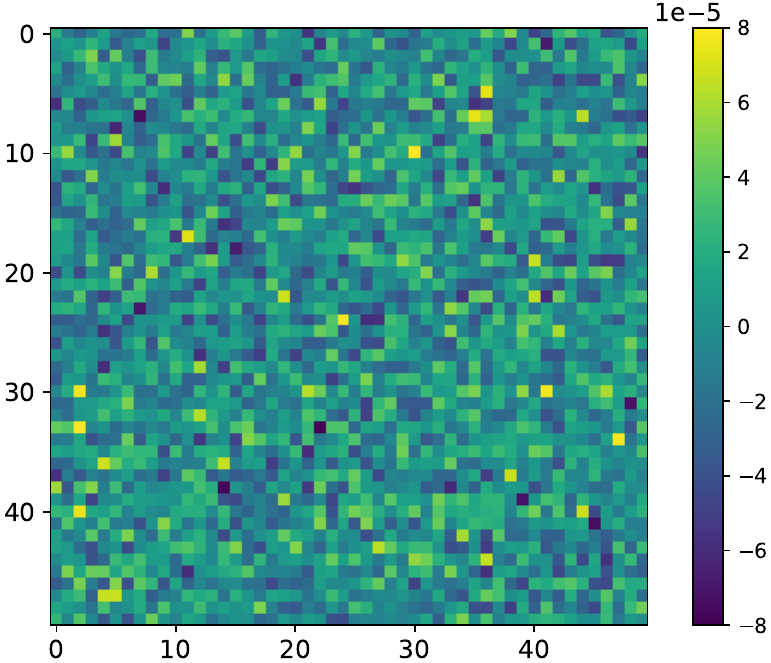}}}
\subfigure[$\eta=0.5$]{
\resizebox{0.32\linewidth}{!}{
\includegraphics[width=2.5in]{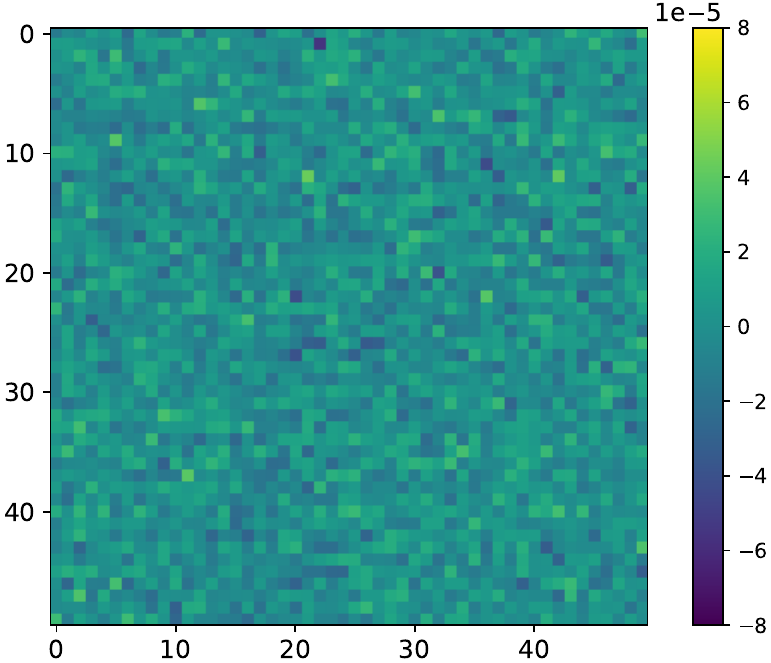}}}
\DeclareGraphicsExtensions.
\caption{Visualization of safety vector (a) and low-rank safety principal components (b, c) with different $\eta$ at the layer model.layers.25.self\_attn.v\_proj, where $\alpha=1$. Visual representations of 2500 random sample positions are provided.}
\label{app-fig:layer-25-v}
\end{figure*}

To analyze the relationship between the safety vector and low-rank safety principal components, we visualized these elements. 
Figures \ref{fig:layer-5-down}, \ref{app-fig:layer-15-v}, and \ref{app-fig:layer-25-v} illustrate as the singular value entropy decreases, the rank of low-rank safety principal components in different layers also decreases, resulting in progressively smoother images. This result is consistent with our preliminary analysis.
Comparing Figures \ref{app-fig:layer-15-v} and \ref{app-fig:layer-15-q}, we note that due to the higher density of safety information encoding in \textit{model.layers.15.self\_attn.q\_proj}, its low-rank safety components exhibit a lower rank when extracting an equivalent proportion of information, producing smoother images. This suggests that, for matrices with high encoding density, a lower rank can effectively represent the safety information corresponding to the safety vector.

\end{document}